\documentclass[aps,prb,amsmath,amssymb,amsfonts,showpacs,superscriptaddress,floatfix]{revtex4} 
\pdfoutput=1
\usepackage{amsmath,amsthm,amsfonts,amssymb}
\usepackage{graphicx}

\usepackage{color}
\numberwithin{equation}{section}
\numberwithin{figure}{section}

\newtheorem{theorem}{Theorem}
\newtheorem{corollary}{Corollary}
\newtheorem{lemma}{Lemma}

\newtheorem{definition}{Definition}

\DeclareMathAlphabet{\mathpzc}{OT1}{pzc}{m}{it}

\newcommand{\be}{\begin{equation}}
\newcommand{\ee}{\end{equation}}

\newcommand{\cT}{{\cal T}}
\newcommand{\rnk}{{\rm rank}}
\newcommand{\cN}{{\cal N}}
\newcommand{\scN}{{\cal N}_c}
\newcommand{\cO}{{\cal O}}
\begin{document}

\title{The Asymptotics of Quantum Max-Flow Min-Cut}

\affiliation{Station Q, Microsoft Research, Santa Barbara, CA 93106-6105, USA}
\affiliation{Quantum Architectures and Computation Group, Microsoft Research, Redmond, WA 98052, USA}
\author{Matthew B.~Hastings}

\begin{abstract}
The quantum max-flow min-cut conjecture relates the rank of a tensor network to the minimum cut in the case that all tensors in the network are identical\cite{mfmc1}.  This conjecture was shown to be
false in Ref.~\onlinecite{mfmc2} by an explicit counter-example.  Here, we show that the conjecture is almost true, in that the ratio of the quantum max-flow to the 
quantum min-cut converges to $1$ as the dimension $N$ of the degrees of freedom on the edges of the network tends to infinity.
The proof is based on estimating moments of the singular values of the network.  We introduce a generalization of ``rainbow diagrams"\cite{rainbow} to tensor networks to estimate the dominant diagrams.  A direct comparison of second and fourth moments lower bounds the ratio of  the quantum max-flow to the 
quantum min-cut by a constant.  To show the tighter bound that the ratio tends to $1$, we consider higher moments.
In addition, we show that the limiting moments as $N \rightarrow \infty$
agree with that in a different ensemble where tensors in the network are chosen independently; this is used to show that the distributions of singular values in the two different ensembles weakly converge to the same limiting distribution.  We present also a numerical study of one particular tensor network, which shows a surprising dependence of the rank deficit on $N \mod 4$ and suggests further conjecture on the limiting behavior of the rank.
\end{abstract}
\maketitle
\section{Introduction}

The quantum max-flow min-cut conjecture was introduced in Ref.~\onlinecite{mfmc1}.  This conjecture relates the rank of a tensor network for a generic choice of tensor to a maximal classical flow (or minimal cut) on the graph corresponding to
the tensor network.
In Ref.~\onlinecite{mfmc2}, various forms of the quantum max-flow min-cut conjecture were considered, and the conjectures were in fact shown to be false.
Here, we consider a particular version of the conjecture, called version 2 in that paper.  Even though the conjecture is not true, we show that the ratio of the actual rank to the rank
predicted by the conjecture converges to $1$
in the limit of large dimension $N$ of the degrees of freedom on the edges of the network.
The proof is statistical in nature, relying on a random choice of tensor in the network in a particular Gaussian ensemble.

We begin by reviewing that particular form of the conjecture (in fact, we consider a special case of the conjecture in that paper, in which all the vertices in the
graph have the same degree and all edges have the same capacity; however, our results can be fairly straightfowardly extended to the general case of the conjecture).  We slightly modify the notation in that paper.

Consider a tensor network.  The tensor network is defined by several pieces of data.  First, there is a graph $G$, with some open edges (i.e., edges that
attach to only one vertex inside the graph) and some closed edges (edges which attach to two vertices).  Second, for each edge, there is an integer, called the capacity in Ref.~\onlinecite{mfmc2}.  Finally, for each vertex, there is a tensor; the number of indices of the tensor is equal to the degree of that vertex and each
index of the tensor corresponds to a distinct edge attached to that vertex; each index ranges over a number of possible values equal to the capacity of that edge.  The entries of the tensor are complex numbers (one could also consider the case that they are real numbers; this would require some modifications of the techniques here).

By contracting the tensor network, the tensor network assigns a complex number for each choice of indices on the open edges.
We partition the open edges into two sets, called the input set and output set.
Define $D_S$ to equal the product of the capacities of all input edges and define $D_T$ to equal the product of the capacities of all
output edges.
This contraction of the tensor network defines a linear operator $L$ from a complex vector space of dimension $D_S$  to one of dimension $D_T$.

For a given tensor network, the rank of this linear operator is termed the quantum flow of the network.  
We define two sets, $S$ and $T$; these sets are the open ends of the input and output open edges, respectively.  We let $V$ be the set of vertices in the graph, not including $S$ and $T$.  We let $\overline V=S \cup T \cup V$; below when we refer to vertices, we only mean vertices in $V$.
A cut is a partition of $\overline V$ into $\overline S \cup \overline T$, where $S \subset \overline S$ and $T \subset \overline T$.
The cut set of the cut is the set of edges $(v,w)$ with $v \in \overline S$ and $w \in \overline T$.
For a given graph, and a given cut set
separating $S$ from $T$, define $D_C$ to equal the product of the capacities of the edges in the cut set.  Then, define the quantum min-cut of the network
to be the minimum of $D_C$ over all cuts.

For a given graph and given capacities, for any choice of tensors, the quantum flow is bounded by the quantum min-cut.  We briefly sketch the proof\cite{mfmc2}.  Given a cut set $C$, cutting the edges in the cut set separates the graph into two graphs, and similarly the tensor network can be split into two tensor networks, one with input $S$ and output $C$ and one with input $C$ and output $T$.  Letting $L_1$ and $L_2$ be the linear operators defined by these networks,
the linear operator $L$ can be written as a product $L=L_2 L_1$.  Clearly, since $L_1$ is a map to a space of dimension $D_C$, it has rank at most $D_C$ and hence so does $L$.

This leaves the question: under what circumstances will the quantum flow equal the quantum min-cut?  Suppose that all edges have the same capacity.  We denote this capacity by $N$ ($c$ was used in Ref.~\onlinecite{mfmc2}; here is one place where we change notation to be more suggestive of ``large $N$" limits in physics).  Then, suppose that all vertices have the same degree $d$.  Choose a single tensor $\cT$ which has $d$ indices, each ranging from $1$ to $N$.  We then use the {\it same} tensor $\cT$ on all vertices of the graph.  For each vertex, choose some ordering assigning indices of the tensor to edges attached to that vertex.  Then, for a given $d$, $N$, graph, and choice of ordering, define the ``quantum max-flow" to be the maximum of the rank over all tensors $\cT$.

Then, the second version of the quantum max-flow min-cut conjecture is that the quantum max-flow is equal to the quantum min-cut.  
This conjecture was shown to be false.

In fact, the conjecture considered in Ref.~\onlinecite{mfmc2} is slightly more general, as edges are allowed to have different capacities.   Then any two vertices which have the same degree and which have the same sequence of capacities of edges attached to the vertex (with the sequence of edges ordered by the assignment of edges to indices of the tensor) are said to have the same ``valence type" and any two vertices with the same valence type
have the same tensor.  Our results can be extended to this case.

Our main result is that the conjecture is ``asynptotically true", in that as $N$ becomes large, the ratio of the quantum max-flow to the quantum min-cut converges to $1$.  Write $QMC(G,N)$ to denote the quantum min-cut for a given graph $G$ and given capacity $N$.  Write $QMF(G,N,O)$ to denote the quantum max-flow for a given graph $G$ and ordering $O$ (the symbol $L$ was used for the ordering in Ref.~\onlinecite{mfmc2} but we use $O$ instead to avoid confusion with $L$ for the linear operator).
For a graph $G$, let $MC(G)$ denote the minimum cut of $G$, where each edge is assigned capacity $1$ to determine the min cut.  Then,
\be
\label{qmcmc}
QMC(G,N)=N^{MC(G)}.
\ee

We show that
\begin{theorem}
\label{strongth}
For all $G,O$,
\be
QMF(G,N,O)=QMC(G,N) \cdot (1-o(1)).
\ee
\end{theorem}
Here, we use a big-O notation where we consider asymptotic behavior as a function of $N$.  The constant factors hidden by the big-O notation may depend on $G,O$.
Of course, the fact that the flow is bounded by the quantum min-cut implies that $QMF(G,N,O)\leq QMC(G,N)$.  The new result will be to
lower bound $QMF(G,N,O)$.

Before proving theorem \ref{strongth}, we will prove a weaker theorem:
\begin{theorem}
\label{mainth}
For all $G,O$,
\be
QMF(G,N,O)=\Theta(QMC(G,N)).
\ee
\end{theorem}
The proof of this will rely on estimating the expectation value of ${\rm tr}(L^\dagger L)$ and ${\rm tr}\Bigl( (L^\dagger L)^2 \Bigr)$, and using a relation between moments of
an operator and its rank.
Most of the work of the paper will be developing tools to estimate moments.

\begin{figure}
\includegraphics[width=0.5in]{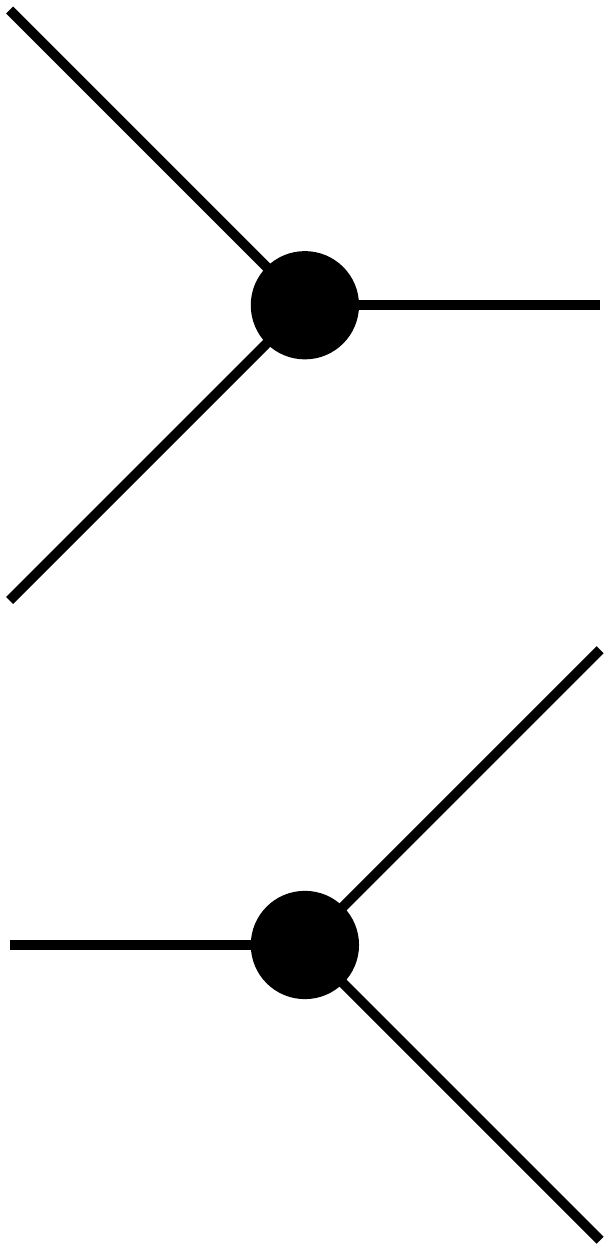}
\caption{Example of a tensor network that is connected in the sense of definition \ref{defconn}.  This network has $|V|=2$ vertices, labelled by solid circles, and $|E|=6$ edges, all of which are open.  There are $|S|=3$ input edges, shown at the left of the figure and $|T|=3$ output edges, shown at the right of the figure.}
\label{figconn}
\end{figure}

Before stating the next theorem about moments, we make some definitions:
\begin{definition}
Given two tensor networks, $\cN_1,\cN_2$ with corresponding graphs $G_1,G_2$, we define their product to be a tensor network with graph $G$ which is the union of $G_1,G_2$.  We write the product as a network $\cN_1 \cdot \cN_2$.  The capacity of an edge in $G$ is given by the capacity of the corresponding edge in $G_1$ or $G_2$ and the ordering of indices of a vertex in $G$ is given by the ordering of the
indices of the corresponding vertex in $G_1$ or $G_2$. 
If $\cN_1,\cN_2$ correspond to linear operators $L_1,L_2$ respectively, then their product corresponds to linear operator $L_1 \otimes L_2$; if $\cN_1,\cN_2$ both have no open edges, so that the corresponding linear operators are scalars, then the linear operator corresponding to the product is simply the product of these scalars.
\end{definition}
\begin{definition}
\label{defconn}
We say that a tensor network $\cN$ with open edges is ``connected" if every vertex is connected to some open edge (either input or output) by a path in the graph corresponding to that
network.
\end{definition}
Note that a tensor network may be connected and yet the graph corresponding to that network may be disconnected.  See Fig.~\ref{figconn}.

Suppose a tensor network $\cN$ is not connected, with $\cN$ corresponding to graph $G$ and linear operator $L$.  Then, let $W$ be the set of vertices which are not connected to an input or output edge.
Then, we can write $\cN$ as a the product of two networks, $\cN_1,\cN_2$, with linear operators $L_1,L_2$ and graphs $G_1,G_2$, where the vertices in $\cN_2$ correspond to the vertices in $W$ and where $\cN_1$ is connected.  Then, $L_2$ is a scalar which is nonzero for generic choice of $\cT$, so that $\rnk(L)=\rnk(L_1)$.
Further, $MC(G)=MC(G_1)$.  So, to prove results about the rank of $L$ in terms of $MC(G)$, we can assume, without loss of generality, that $\cN$ is connected.
So, unless explicitly stated otherwise, all tensor networks with open edges that we consider will be connected and all linear operators $L$ will correspond to connected tensor networks.

In general, we use $|E|$ to denote the number of edges of a tensor network (including open edges) and use $|V|$ to denote the number of vertices.For notational simplicity, throughout the paper we
label closed edges by the pair of vertices $(v,w)$ to which they are attached; the results also apply to the case in which there are multiple edges
connecting a pair of vertices in which case an additional label must be given to the edge to indicate for which edge it is; for notational simplicity, we
do not write this extra label.
We will later define a Gaussian ensemble to choose the tensors.  For most of the paper, we consider the case that all the tensors in
the network are identical, given by some fixed tensor $\cT$ drawn
from this ensemble.  However, we sometimes also consider the case that the tensors in the network are not identical, and instead are chosen independently
from the Gaussian ensemble.  If we need to distinguish these two cases, we refer to them as the {\it identical ensemble} and {\it independent}
ensemble.  If not otherwise stated, we are referring to the indentical ensemble.  We use $E[\ldots]$ to denote an expectation value in the identical ensemle
and $E_{\rm ind}[\ldots]$ to denote an expectation value in the independent ensemble.

We can now state the following theorem proven later.
\begin{theorem}
\label{mainth2}
Consider a tensor network $\cN$ (assumed connected as explained above) with corresponding graph $G$.
Then for the linear operator $L$ corresponding to the tensor network,
for any $k \geq 1$, we have
\be
\label{main2}
E[{\rm tr}\Bigl((L^\dagger L)^k\Bigr)]=c(G,k) \cdot N^{k |E| - (k-1) MC(G)}+\cO(N^{k |E| - (k-1) MC(G)-1}),
\ee
where 
where $c(G,k)$ denotes a positive integer that depends upon the graph $G$ and on $k$ (the constant $c(G,k)$ does not depend on the ordering $O$).
Further,
\be
\label{main2ind}
E_{\rm ind}[{\rm tr}\Bigl((L^\dagger L)^k\Bigr)]=c(G,k) \cdot N^{k |E| - (k-1) MC(G)}+\cO(N^{k |E| - (k-1) MC(G)-1}),
\ee
with the same constant $c(G,k)$ in Eq.~(\ref{main2}) as in Eq.~(\ref{main2ind}).

Further, for any graph $G$, the constant $c(G,k)$is bounded by an exponential in $k$, i.e., $C(G,k)\leq c_1 \cdot c_2^k$, where the constants $c_1,c_2$ depend upon $G$.
\end{theorem}

This theorem requires that the network be connected.  As a trivial example to show how  this is necessary, consider a network which computes a scalar.  Suppose that the tensors in the network all are also scalars; i.e., the vertices have degree $0$.  In this trivial example, suppose that $|V|=2$.
This network is {\it not} connected.
If the two ``tensors" at the two different vertices are the scalars $x,y$, then the operator $L$ is equal to the scalar $xy$.  If one picks $x,y$ independent complex Gaussians with probability distribution function
$\frac{1}{\pi}\exp(-|z|^2)$,
then one can readily compute $E_{\rm ind}[{\rm tr}(L^\dagger L)]=E_{\rm ind}[|x|^2 |y|^2]=1$.
However, if we choose $x$ Gaussian and set  $y=x$, then $E[|x|^2 |y|^2]=E[|x|^4]=2$ and so the expectation values would be different in the
independent and identical ensembles.  
This example might seem strange (having degree $0$), so the reader can also consider, for example, a tensor network with $2$ vertices and $d$ edges, with no open edges so that
all edges connect one vertex to the other; using the techniques later to evaluate expectation values, the reader can check that the expectation values will differ.

Thus, if we define $$K=\Bigl( N^{|E|-MC(G)} \Bigr)^{-1/2} L$$ and define $${\rm av}(\ldots)=\frac{1}{N^{MC(G)}} {\rm tr}(\ldots),$$ we have
\be
E[ {\rm av}\Bigl( (K^\dagger K)^k \Bigr)]= c(G,k)+\cO(1/N),
\ee
and
\be
E_{\rm ind}[ {\rm av}\Bigl( (K^\dagger K)^k \Bigr)]
= c(G,k)+\cO(1/N).
\ee

The notation ${\rm av}(\ldots)$ is intended to be suggestive as follows: we know\cite{mfmc2} that if the tensors are chosen independently for generic choice of tensors then
$L$ (and hence $K$) has rank $N^{MC(G)}$.  Hence, for given $K$, the expectation value, of the $2k$-th moment of a randomly chosen non-zero singular
value of a random tensor in the independent ensemble is $E_{\rm ind}[{\rm av}(\Bigl( (K^\dagger K)^k \Bigr)]$.  Let $\mu^{\rm ind}_N$ be the distribution function of a randomly chosen singular value for a randomly
chosen tensor.
By the fact that $c(G,k)$ are bounded by an exponential in $k$, the distributions $\mu^{\rm ind}_N$ converge weakly to a limit\cite{carleman}.
Suppose instead we choose the tensors all identically from the Gaussian ensemble and then randomly choose one of the largest $N^{MC(G)}$ singular values (i.e., if there are $\rnk(L) \leq N^{MC(G)}$ non-zero singular values, then with probability $\rnk(L)/N^{MC(G)}$ we choose one of the non-zero singular values and otherwise the result we choose a zero singular value).  Let this distribution be $\mu_N$.  Then, the moments of $\mu_N$ are the same as the moments of $\mu^{\rm ind}_N$ up to $O(1/N)$ corrections and so
we have the further corollary that
\begin{corollary}
\label{weakconv}
$\mu_N$
converges weakly to the same limit as $\mu^{\rm ind}_N$.  The limiting distribution has compact support
\end{corollary}

Also,
\begin{corollary}
\label{cor2}
For all $\epsilon>0$, for all $n>0$, there is a constant $c$ such that for all sufficiently large $N$
the probability that the largest singular value of $L$ is greater than or equal to
$x\sqrt{N^{|E|-MC(G)+\epsilon}}$  is bounded by $c x^{-n}$.
\begin{proof}
Let $\lambda$ be the largest singular value of $L$.  $E[\lambda^{2k}]\leq c(G,k)(1+\cO(1/N)) N^{k |E| - (k-1) MC(G)}$.  So, the probability that
$\lambda\geq x \sqrt{N^{|E|-MC(G)+\epsilon}}$ is 
bounded by
\be
\frac{E[\lambda^{2k}]}{x^{2k} N^{k|E|-kMC(G)+k\epsilon}}=
c(G,k)x^{-2k} (1+\cO(1/N)) N^{MC(G)-k\epsilon},
\ee
and choosing $k>MC(G))/\epsilon$ and $k\geq n/2$, this is bounded by a constant times $x^{-n}$ for all sufficiently large $N$.
\end{proof}
\end{corollary}

Since if the tensors are chosen independently, one has $QMC(G,N)=QMF(G,N)$ generically, one might naively guess that corollary
\ref{weakconv} implies theorem \ref{strongth} as follows: for independent choice of tensors, the linear operator $L$ will generically have $QMC(G,N)$ nonzero
singular values and so one might expect that when the tensors are chosen identically the linear operator $L$
will have nearly $QMC(G,N)$ singular values.  The trouble with this naive argument is that it is conceivable that in the independent ensemble
the linear operator $K$ will have $QMC(G,N)$ nonzero singular values but that with high probability a constant fraction of them will have magnitude which is $o(1)$ so that $\mu^{\rm ind}_N$ will converge to, for example, a sum of a smooth function plus a $\delta$-function at the origin.  So, to prove theorem \ref{strongth} we instead give a more detailed analysis of higher moments.

We begin in section \ref{msec} by lower bounding the rank in terms of traces of  moments of the linear operator $L$.  We also define the appropriate Gaussian ensemble for $\cT$
in this section, and give a combinatorial method for computing these traces for this ensemble.  Then, in section \ref{fsec}, we use these methods to estimate the expectation value of
${\rm tr}(L^\dagger L)$.  In sections \ref{mlbsec},\ref{mubsec}, we show how to estimate expectation values of traces of higher moments of $L^\dagger L$, as well as expectation values
of products of such traces; this is done by combining a lower bound in section \ref{mlbsec} for these expectation values with an upper bound in section \ref{mubsec}.
The techniques for computing the expectation value show that the results are indeed the same, up to $\cO(1/N)$, for the two ensembles.
In section \ref{proofstrongth} we complete the proof of theorem \ref{strongth}.
In section \ref{sectionvar} we collect some results on variance that are not needed elsewhere but may be of independent interest.
In section \ref{sectionnumerics} we present some numerical simulations.

\section{Moments Bound and Definition of Ensemble}
\label{msec}
Let $\rnk(L)$ denote the rank of a linear operator $L$.
We have the following bound:
\begin{lemma}
\label{momentslemma}
For any linear operator $L$, and any integer $k>1$,
\be
\label{moments}
\rnk(L)^{k-1} \geq \frac{{\rm tr}(L^\dagger L)^k}{{\rm tr}\Bigl((L^\dagger L)^k\Bigr)},
\ee
assuming that the denominator of the right-hand side is nonzero.
In the special case $k=2$ used later we have
\be
\rnk(L)\geq \frac{{\rm tr}(L^\dagger L)^2}{{\rm tr}\Bigl((L^\dagger L)^2\Bigr)}.
\ee
\begin{proof}
Let the non-zero singular values of $L$ be $\lambda_1,...,\lambda_{\rnk(L)}$.
Let the vector $v$ be $(\lambda_1^2,...,\lambda_{\rnk(L)}^2)$.  Let the vector $w$ be $(1,1,...,1)$.
Then, by H\"{o}lder's inequality applied to vectors $v,w$,
\be
\sum_{i=1}^{\rnk(L)} \lambda_i^2 \leq \Bigl( \sum_{i=1}^{\rnk(L)}  \lambda_i^{2p} \Bigr)^{1/p} \rnk(L)^{1/q}
\ee
for any $p,q$ with $1/p+1/q=1$.  Choosing $p=k$, $q=k/(k-1)$, and raising the above equation to the $k$-th power, we get
\be
{\rm tr}(L^\dagger L)^k\leq  {\rm tr} \Bigl((L^\dagger L)^k\Bigr)  \rnk(L)^{k-1},
\ee
as claimed
(in the special case of $k=2$, we can use Cauchy-Schwarz instead of H\"{o}lder).
\end{proof}
\end{lemma}

We will prove Theorem \ref{mainth} by estimating the expected value of the numerator and denominator of Eq.~(\ref{moments}) for a particular random
ensemble of tensors.  The ensemble that we choose is that the entries of the tensor $\cT$ will be chosen independently and indentically distributed, using
a Gaussian distribution with probability density
$$\frac{1}{\pi}\exp(-|z|^2)$$ so that $E[|z|^2]=1$.

We estimate the expectation value of the numerator and denominator of Eq.~(\ref{moments}) independently, rather than estimating the expectation value of the ratio, and use the following lemma:
\begin{lemma}
\label{exists}
Let $E[{\rm tr}(L^\dagger L)^k]$ and $E[{\rm tr}\Bigl((L^\dagger L)^k\Bigr)]$ be given and nonzero for $k>1$. 
Then there must exist some tensor $\cT_0$ for which the corresponding linear operator $L_0$ obeys
\be
\rnk(L_0)^{k-1} \geq \frac{E[{\rm tr}(L^\dagger L)^k]}{E[{\rm tr}\Bigl((L^\dagger L)^k\Bigr)]}.
\ee
Further, since $E[{\rm tr}(L^\dagger L)^k] \geq E[{\rm tr}(L^\dagger L)]^k$,
\be
\rnk(L_0)^{k-1} \geq \frac{E[{\rm tr}(L^\dagger L)]^k}{E[{\rm tr}\Bigl((L^\dagger L)^k\Bigr)]}.
\ee
\begin{proof}
Given $E[{\rm tr}(L^\dagger L)^k]$ and $E[{\rm tr}\Bigl((L^\dagger L)^k\Bigr)]$, there must exist some $\cT_0$ for which
\be
\frac{{\rm tr}(L_0^\dagger L_0)^k}{{\rm tr}\Bigl((L_0^\dagger L_0)^k\Bigr)}
 \geq
\frac{E[{\rm tr}(L^\dagger L)]^k}{E[{\rm tr}\Bigl((L^\dagger L)^k\Bigr)]}.
\ee
The result then follows from Eq.~(\ref{moments}).
\end{proof}
\end{lemma}

\begin{figure}
\includegraphics[width=1.0in]{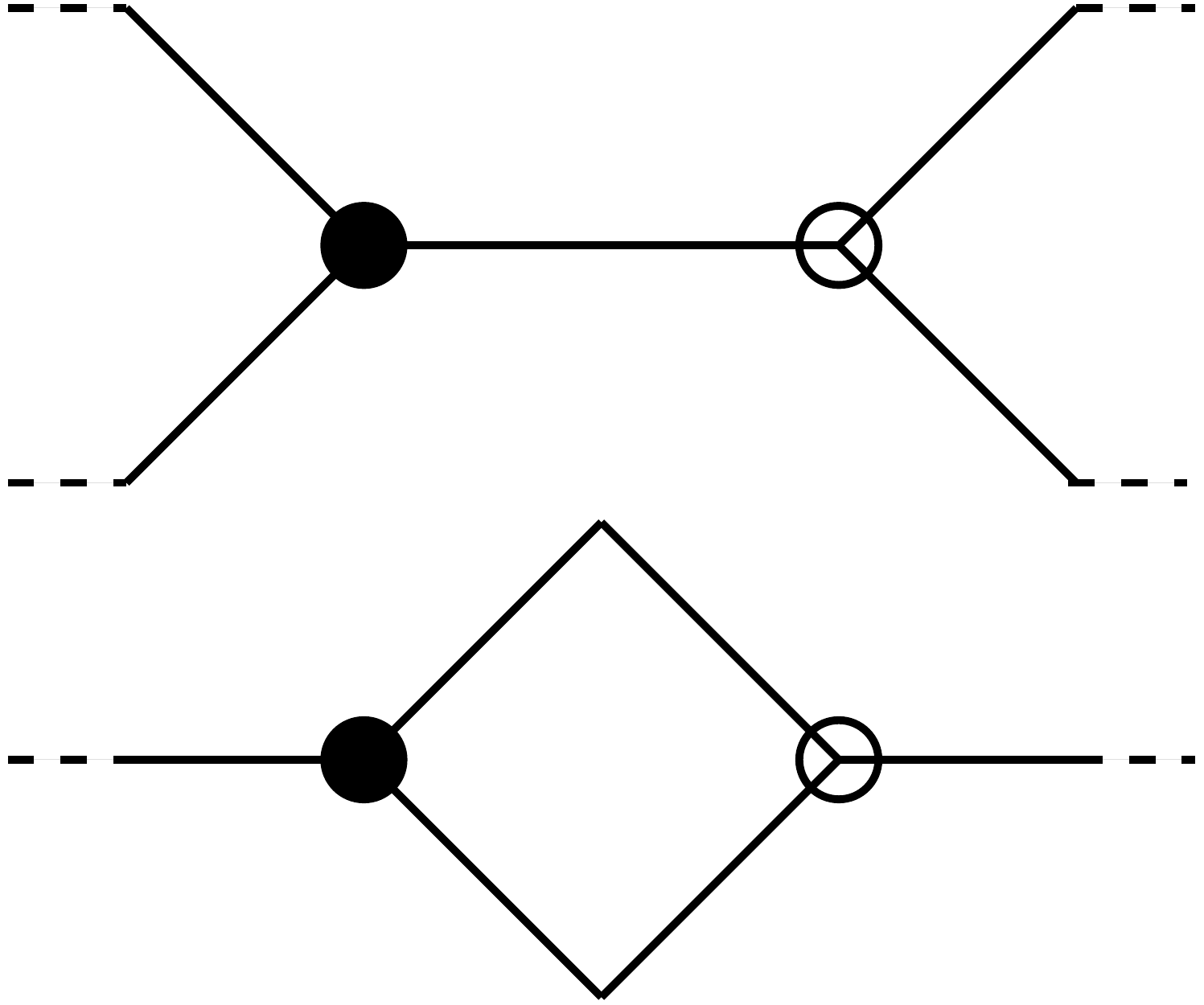}
\caption{Closed tensor network for computing ${\rm tr}(L^\dagger L)$ for the linear operator corresponding to the tensor network shown in Fig.~\ref{figconn}.  Closed circles represent tensors $\cT$, while open circles represent tensors $\overline \cT$.  The dashed lines on the $4$ open edges are used to indicate that the edges on the right-hand side of the figure
should be joined to the edges on the left-hand side of the figure so that the tensor network is closed, computing the trace.  Without these dashed lines, the tensor network
would have $4$ output edges and would compute the linear operator $L^\dagger L$ with the four output edges on the right and the four input edges on the left (while a mathematical expression such as $L^\dagger L$ has its ``input on the right" and its ``output on the left", tensor networks are conventionally read from left to right instead.}
\label{figLdaggerL}
\end{figure}

\begin{figure}
\includegraphics[width=2.0in]{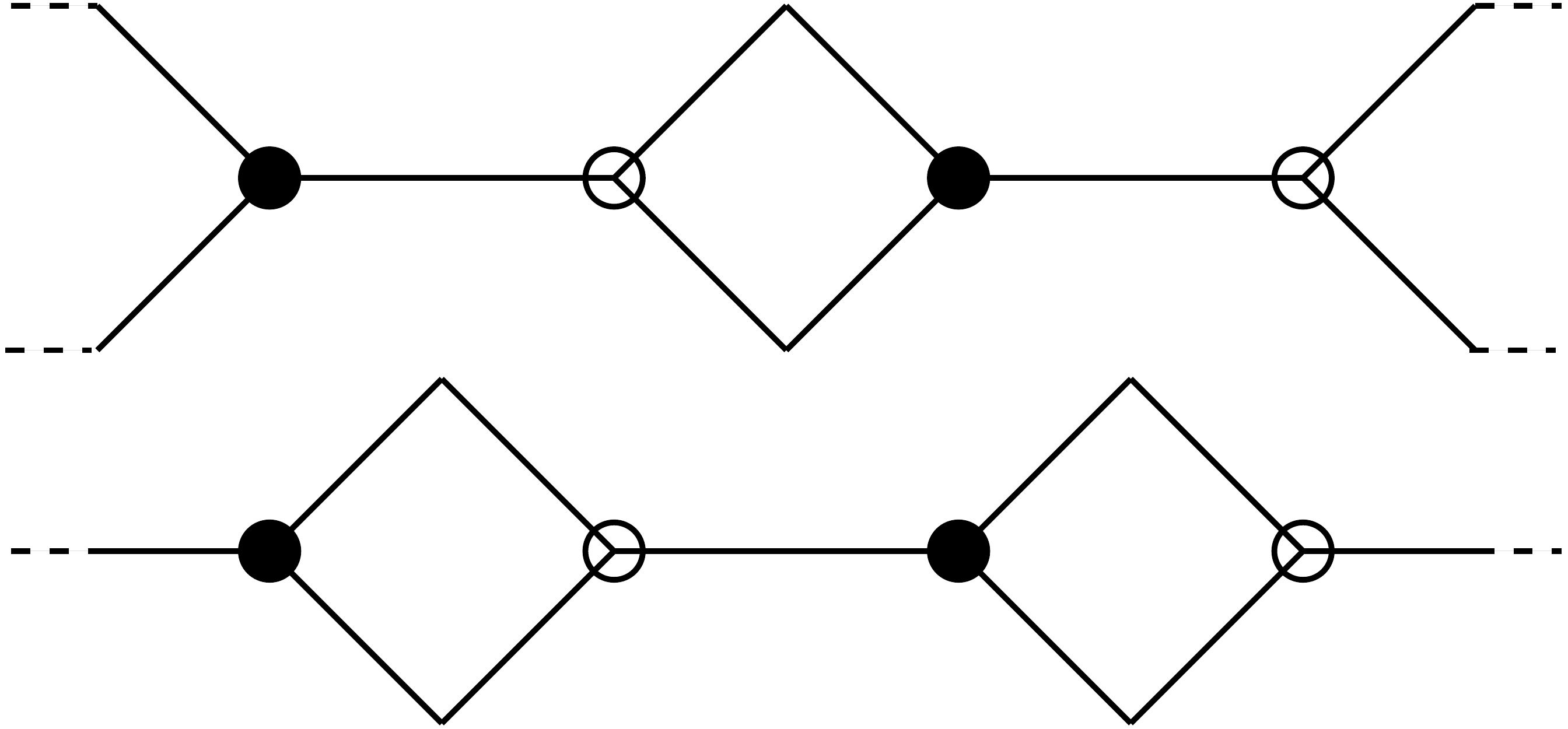}
\caption{Closed tensor network for computing ${\rm tr}\Bigl((L^\dagger L)^2\Bigr)$ for the linear operator corresponding to the tensor network shown in Fig.~\ref{figconn}.}
\label{figLdaggerL2}
\end{figure}

For a given tensor $\cT$, traces 
such as ${\rm tr}(L^\dagger L)$ and ${\rm tr}(L^\dagger L L^\dagger L)$
are also given by tensor networks; they are tensor
networks with no open edges so that their contraction yields a scalar, and this scalar is equal to the desired trace.
\begin{definition}
We refer to such networks with no
open edges as {\it closed tensor networks}.
\end{definition}
See Figs.~\ref{figLdaggerL},\ref{figLdaggerL2}.  The notation in Fig.~\ref{figLdaggerL} with closed and open circles to denote $\cT$ and $\overline \cT$ and dashed lines to
denote edges that should be joined to compute a trace will be used in figures from here on.
We use $\cN$ to denote the network with open edges that is used to define $L$ and we use $\scN$ to denote various different tensor
networks with no open edges; the networks $\scN$ that we consider will correspond to traces such as ${\rm tr}(L^\dagger L)$, ${\rm tr}(L^\dagger L L^\dagger L)$, and so on; so, $\scN$ is derived from $\cN$ and from the choice of the particular trace.

\begin{definition}
We introduce notation: $\scN({\rm tr}(L^\dagger L))$ indicates the closed tensor network corresponding to ${\rm tr}(L^\dagger L)$, while $\scN({\rm tr}(L^\dagger L L^\dagger L))$ indicates the tensor network corresponding to ${\rm tr}(L^\dagger L L^\dagger L)$, and so on.
Additionally, we consider closed tensor networks which correspond to products of traces, so that
$\scN({\rm tr}(L^\dagger L) {\rm tr}(L^\dagger L L^\dagger L))$ indicates the tensor network corresponding to ${\rm tr}(L^\dagger L) {\rm tr}(L^\dagger L L^\dagger L)$. 
Such a tensor network $\scN({\rm tr}(L^\dagger L) {\rm tr}(L^\dagger L L^\dagger L))$ is the product of the tensor networks
$\scN({\rm tr}(L^\dagger L))$ and $\scN({\rm tr}(L^\dagger L L^\dagger L))$.
\end{definition}

We now explain how to compute the expectation value of the contraction of a closed tensor network $\scN$; for brevity, we will simply refer to this as
``the expectation value of the tensor network", rather than ``the expectation value of the contraction of the tensor network".
The tensor network $\scN$ is a polynomial in the entries of $\cT$ and $\overline \cT$, where the overline denotes the complex conjugate.  Hence, we can use Wick's theorem to compute the expectation value.  Suppose that the tensor network has $M$ different tensors $\cT$ and $M'$ different tensors $\overline \cT$.
The expectation value is nonzero only if $M=M'$.
Wick's theorem computes the expectation value by summing over all possible pairings of a tensor $\cT$ with a tensor $\overline \cT$, so that every $\cT$ is paired with a unique $\overline \cT$; that is, there are $M!$ different pairings to sum over.  
Given a tensor $\cT$ with $d$ indices, with entries of the tensor written $\cT_{i_1,...,i_d}$, we have
\be
E[\cT_{i_1,...,i_d} \overline\cT_{j_1,...,j_d}]=\delta_{i_1,j_1} ... \delta_{i_d,j_d}.
\ee

\begin{figure}
\includegraphics[width=1.0in]{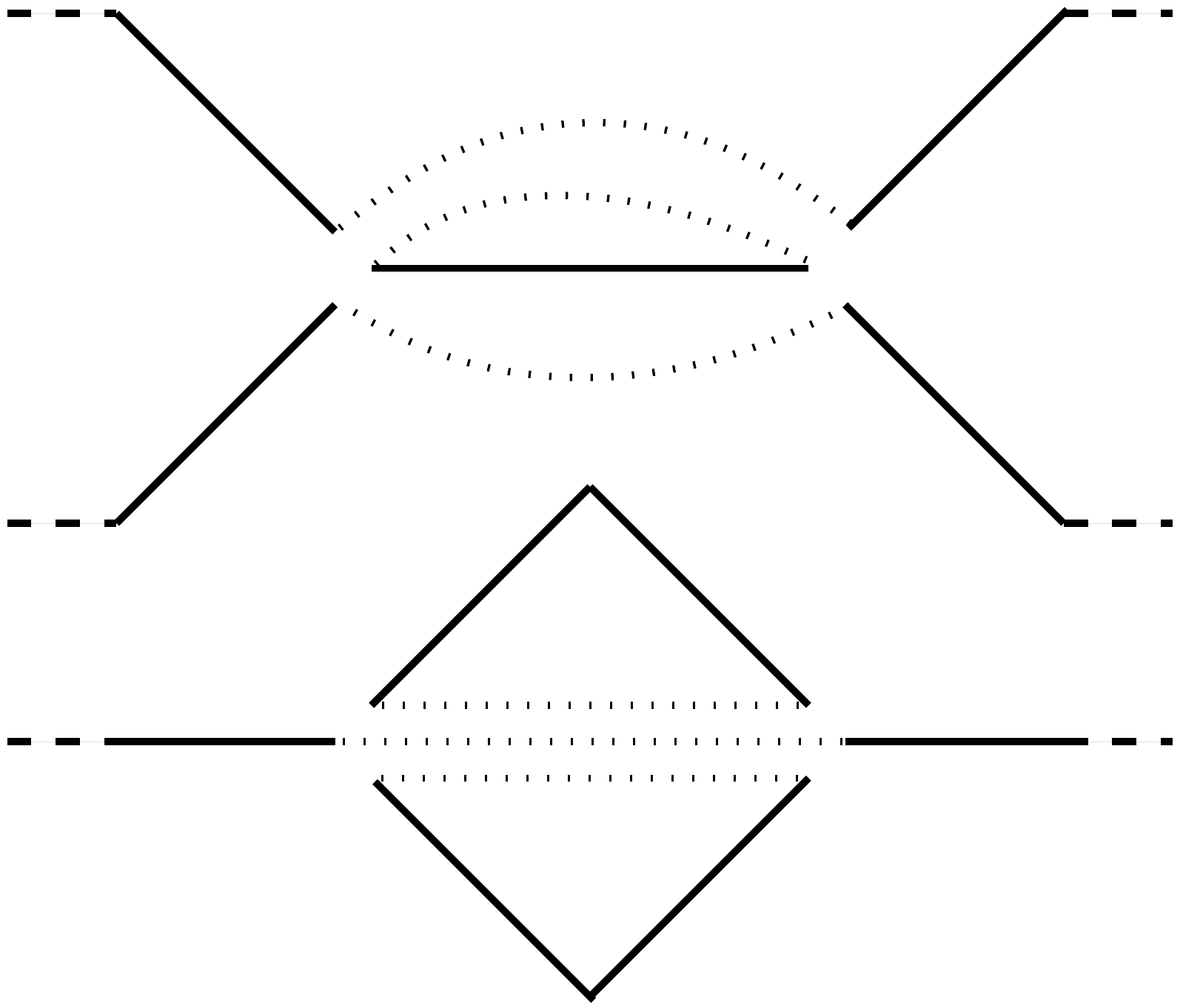}
\caption{A possible pairing of the closed tensor network in Fig.~\ref{figLdaggerL}.  Dotted lines are used to connect edges which end at vertices which are paired; each vertex in a pair has $d=3$ different edges: we connected the edges which have the same index for the local ordering.  This pairing has $6$ different closed loops.  There is one other possible pairing for this network; that pairing would have only $3$ closed loops.}
\label{figloops}
\end{figure}

These $\delta$-functions can be represented graphically as follows.
For each pairing, we define a new graph by removing every vertex (leaving all edges with both ends open) and then
for each pair of vertices $v,w$ which are paired with each other, for each $a \in \{1,...,d\}$, we attach the end of the edge which corresponded to the $a$-th index of the tensor at vertex $v$ to the
end of the edge which corresponded to the $a$-th index of the tensor at vertedx $w$. 
Then, the tensor network breaks into a set of disconnected closed loops.  See Fig.~\ref{figloops}.
\begin{definition}
\label{cldef}
A closed loop created by a pairing consists of a sequence of edges $(v_1,w_1),(v_2,w_2),...,(v_l,w_l)$; no repetition of edges is allowed in a closed loop, but vertices
may be repeated.  
The closed loops created by the pairing are such that $w_i$ is paired with $v_{i+1}$ (identiyfing $v_{l+1}$ with $v_1$) with the local ordering assigning the same index of the
tensor to the edge $(v_i,w_i)$  at $w_i$ as it assigns to edge $(v_{i+1},w_{i+1})$ at $v_{i+1}$.
The choice of starting edge in the sequence is irrelevant as is the direction in which the edges are traversed; i.e., the
sequences  $(v_1,w_1),(v_2,w_2),...,(v_l,w_l)$ and $(v_2,w_2),\ldots,(v_l,w_l),(v_1,w_1)$ and $(w_l,v_l),\ldots,(w_2,v_2),(w_1,v_1)$ all denote
the same closed loop.
\end{definition}
Suppose for a given pairing $\pi$ that the number of closed loops is
equal to $C(\pi)$.  Then, the sum over indices on the edges for the given pairing is $N^{C(\pi)}$.
Thus, for a tensor network $\scN$ with $M=M'$, the expectation value is equal to
\be
\label{value}
E[{\scN}]=\sum_{\pi \, {\rm pairing} \, \cT \, {\rm with} \, \overline \cT} N^{C(\pi)}.
\ee

Note that every term in Eq.~(\ref{value}) is positive.  So, once we have found that there exists some pairing $\pi_0$ with some given $C(\pi_0)$, we have
established that $E[{\scN}] \geq N^{C(\pi_0)}$.  Further, the pairing or pairings with the largest $C(\pi_0)$ give the dominant contribution in the
large $N$ limit.
We define
\be
C_{\rm max}={\rm max}_{\pi} C(\pi),
\ee
and define $n_{\rm max}$ to be the number of distinct pairings $\pi$ with $C(\pi)=C_{\rm max}$;
then
\be
\label{valueasympt}
E[{\scN}]= \Theta(N^{C_{\rm max}}).
\ee
and
\be
\label{valueasympt2}
E[{\scN}]=\Bigl(1-\cO(1/N)\Bigr) \cdot n_{\rm max} N^{C_{\rm max}}.
\ee

\section{Estimating First Moment}
\label{fsec}
We now estimate $E[{\rm tr}(L^\dagger L)]$.
\begin{lemma}
\label{sqlemma}
Given a tensor network $\cN$ obtained from a graph $G$ with $|V|$ vertices and $|E|$ edges, with corresponding linear operator $L$,
for $\scN({\rm tr}(L^\dagger L))$ we have
\be
C_{\rm max}=|E|,
\ee
and
\be
n_{\rm max}=1.
\ee
\begin{proof}
First, we explicitly give a pairing $\pi$ for which $C(\pi)=|E|$.  Note that the number of vertices in the graph $G'$ for tensor network $\scN({\rm tr}(L^\dagger L))$
equals $2|V|$.  There are $|V|$ vertices with tensor $\cT$ and $|V|$ vertices with tensor $\overline \cT$.  There is an obvious pairing $\pi$  of the vertices in
$G'$ exemplified in Fig~\ref{figloops} for one particular network.  For this pairing $\pi$, we have $C(\pi)=|E|$.  We can define this pairing formally for arbitrary $G$ as follows.  Suppose $G$ has vertex set ${\cal V}$, and edge set ${\cal E}$.  Then, $G'$ has vertices labelled by a pair $(v;\sigma)$ where $v\in {\cal V}$ and $\sigma\in \{1,2\}$.  If $\sigma=1$ then the vertex
has tensor $\cT$, and if $\sigma=2$ then the vertex has tensor $\overline \cT$.
If $(v,w)$ is an edge in $G$, then $((v;\sigma),(w;\sigma))$ is an edge of $G'$ for $\sigma\in \{1,2\}$.  Additionally, for every open edge in $G$ there is an
edge in $G'$; for each open edge attached to a vertex $v$, then there is an edge $((v;1),(v;2))$ in $G'$.  There are no other edges in $G'$ other than
those given by these rules.
The ordering of edges attached to vertices in $G'$ is defined in the obvious way: if an edge $e$ in $G$ attached to a vertex $v$ corresponds to the $j$-th index of the tensor, then the edge in $G'$ obtained from $e$ in the above rules attached to $(v;\sigma)$ also corresponds to the $j$-th index of the tensor.
This pairing $\pi$ is then the pairing which pairs each vertex $(v;1)$ with $(v;2)$.
This pairing gives one closed loop for every edge of $G$ so that $C(\pi)=|E|$.

We next show that there is no pairing $\pi$ for which $C(\pi)>|E|$.  A closed loop corresponds to a sequence of edges in $G'$ which we write as $(v_1,w_1),(v_2,w_2),...,(v_l,w_l)$ for some $l$; we say that such a loop ``has $l$ edges".  The pairing is such that $w_i$ is paired with $v_{i+1}$ for $i<l$ and $w_l$ is paired with $v_1$.  If $l=1$, then $v_1$ has a tensor $\cT$ and $w_1$ has a tensor $\overline \cT$ and so this edge in $G'$ is obtained from an open edge in $G$.
If $l>1$, then vertices $v_i$ and $w_i$ for odd $i$ have tensors $\cT$ while for even $i$ they have tensors $\overline \cT$.
So, if $l>1$, then $l$ is even.

Let $n_1$ be the number 
 of closed loops in the pairing with $l=1$; $n_1$ is bounded by the number of open edges in $G$, which is equal to $|S|+|T|$.  The number of closed loops with $l>1$ is then bounded
by $(1/2)(|E'|-n_1)$, where $|E'|$ is the number of edges in $G'$, since every closed loop with $l>1$ must have at least two edges.  Note that $|E'|=2|E|-|S|-|T|$.

So, the number of closed loops is bounded by
\be
\label{maxC}
C(\pi) \leq n_1+(1/2)(|E'|-n_1),
\ee
which is maximal when $n_1$ is as large as possible, i.e., $n_1=|S|+|T|$.  In this case, the
maximum number of closed loops equals $|E|$.
So, $C_{\rm max}=|E|$.

We now show that $n_{\rm max}=1$.  Consider a pairing $\pi$ with $C(\pi)=|E|$ so that $n_1=|S|+|T|$.
Thus, for every open edge in $G$, there must be a closed loop containing just the edge in $G'$ corresponding to that open edge in $G$.
Further, to have $C(\pi)=|E|$, we must have that no loops have more than two edges, as otherwise $C(\pi)<n_1+(1/2)(|E'|-n_1)$.
A loop with two edges corresponds to a sequence $(v_1,w_1),(v_2,w_2)$ with $w_1,v_2$ paired and $v_1,w_2$ paired.
 This then allows us to show that $n_{\rm max}=1$ as follows.
Since for every open edge in $G$, there is a closed loop of length $l=1$ containing just the corresponding edge in $G'$, any vertex $(v;1)$
which is attached to an open edge must be paired with $(v,2)$.  That is, for all vertices $v \in G$ attached to an open edge, we pair $(v;1)$ and $(v;2)$.
Now consider a vertex $w\in G$ which neighbors some vertex $v\in G$ such that we pair $(v;1)$ and $(v;2)$.  Then, there is some
edge $((w;1),(v;1))$ and this edge must be in a closed loop with two edges.  Since we pair $(v;1)$ with $(v;2)$, there must be a loop
containing edges $((w;1),(v;1))$ and $((v;2),((w;2))$.  Then, since this loop must have length $2$, we must pair $(w;1)$ with $(w;2)$.
Let $P$ be the set of vertices $v\in G$ such that we pair $(v;1)$ with $(v;2)$; we have shown that $P$ contains all vertices attached to an open edge and $P$ contains all vertices connected to a vertex in $P$ by an edge.  So, since the network is connected, $P$ must contain all vertices and so there is indeed only one such pairing.
\end{proof}
\end{lemma}

\section{Lower Bound For Higher Moments}
\label{mlbsec}
We now lower bound $E[{\rm tr}(L^\dagger L L^\dagger L)]$ and other higher moments.
First let $G'$ denote the graph for tensor network $\scN({\rm tr}(L^\dagger L L^\dagger L))$.  Let us formally define $G'$, in a way similar to that in which a different graph (also called $G'$) was defined in lemma \ref{sqlemma}.
 Suppose $G$ has vertex set ${\cal V}$, and edge set ${\cal E}$.  Then, $G'$ has vertices labelled by a pair $(v;\sigma)$ where $v\in {\cal V}$ and $\sigma\in \{1,2,3,4\}$.  If $\sigma\in \{1,3\}$ then the vertex
has tensor $\cT$, and if $\sigma\in \{2,4\}$ then the vertex has tensor $\overline \cT$.
If $(v,w)$ is an edge in $G$, then $((v;\sigma),(w;\sigma))$ is an edge of $G'$ for $\sigma\in \{1,2,3,4\}$.
Additionally, for every open edge in $G$ there is are two
edges in $G'$; for each open edge attached to a vertex $v$, if the edge is an input edge then there are edges $((v;1),(v;2))$ and $((v;3),(v;4))$ in $G'$ while if the edge is an output edge then there are edges $((v;2),(v;3))$ and $((v;4),(v;1))$.  There are no other edges in $G'$ other than
those given by these rules.
The ordering of edges attached to vertices in $G'$ is defined in the obvious way: if an edge $e$ in $G$ attached to a vertex $v$ corresponds to the $j$-th index of the tensor, then the edge in $G'$ obtained from $e$ in the above rules attached to $(v;\sigma)$ also corresponds to the $j$-th index of the tensor.
If we instead consider a higher moment $E[{\rm tr}\Bigl((L^\dagger L)^k\Bigr)]$, we define a graph $G'$ for tensor network
$\scN({\rm tr}\Bigl((L^\dagger L)^k\Bigr))$ similarly.
Now $G'$ has vertices labelled by a pair $(v;\sigma)$ where $v\in {\cal V}$ and $\sigma\in \{1,2,\ldots,2k\}$.  If $\sigma$ is odd then the vertex
has tensor $\cT$, and if $\sigma$ is even then the vertex has tensor $\overline \cT$.
If $(v,w)$ is an edge in $G$, then $((v;\sigma),(w;\sigma))$ is an edge of $G'$ for all $\sigma$.
Additionally, for every open edge in $G$ there is are $k$
edges in $G'$; for each open edge attached to a vertex $v$, if the edge is an input edge then there are edges $((v;\sigma),(v;\sigma+1))$ for $\sigma$ odd
 while if the edge is an output edge then there are edges $((v;\sigma),(v;\sigma+1))$ for $\sigma$ even.
We regard $\sigma$ as periodic mod $2k$, so that $\sigma=2k+1$ is the same
as $\sigma=1$.
The ordering of edges attached to vertices in $G'$ is defined in the obvious way: if an edge $e$ in $G$ attached to a vertex $v$ corresponds to the $j$-th index of the tensor, then the edge in $G'$ obtained from $e$ in the above rules attached to $(v;\sigma)$ also corresponds to the $j$-th index of the tensor.

Thus, for example, for the closed tensor network in Fig.~\ref{figLdaggerL2}, the four vertices in the top row correspond to $\sigma=1,2,3,4$ from {\it left} to {\it right} (recall that the input of the tensor network is on the left), as do the four
vertices in the bottom row.

\begin{figure}
\includegraphics[width=2.0in]{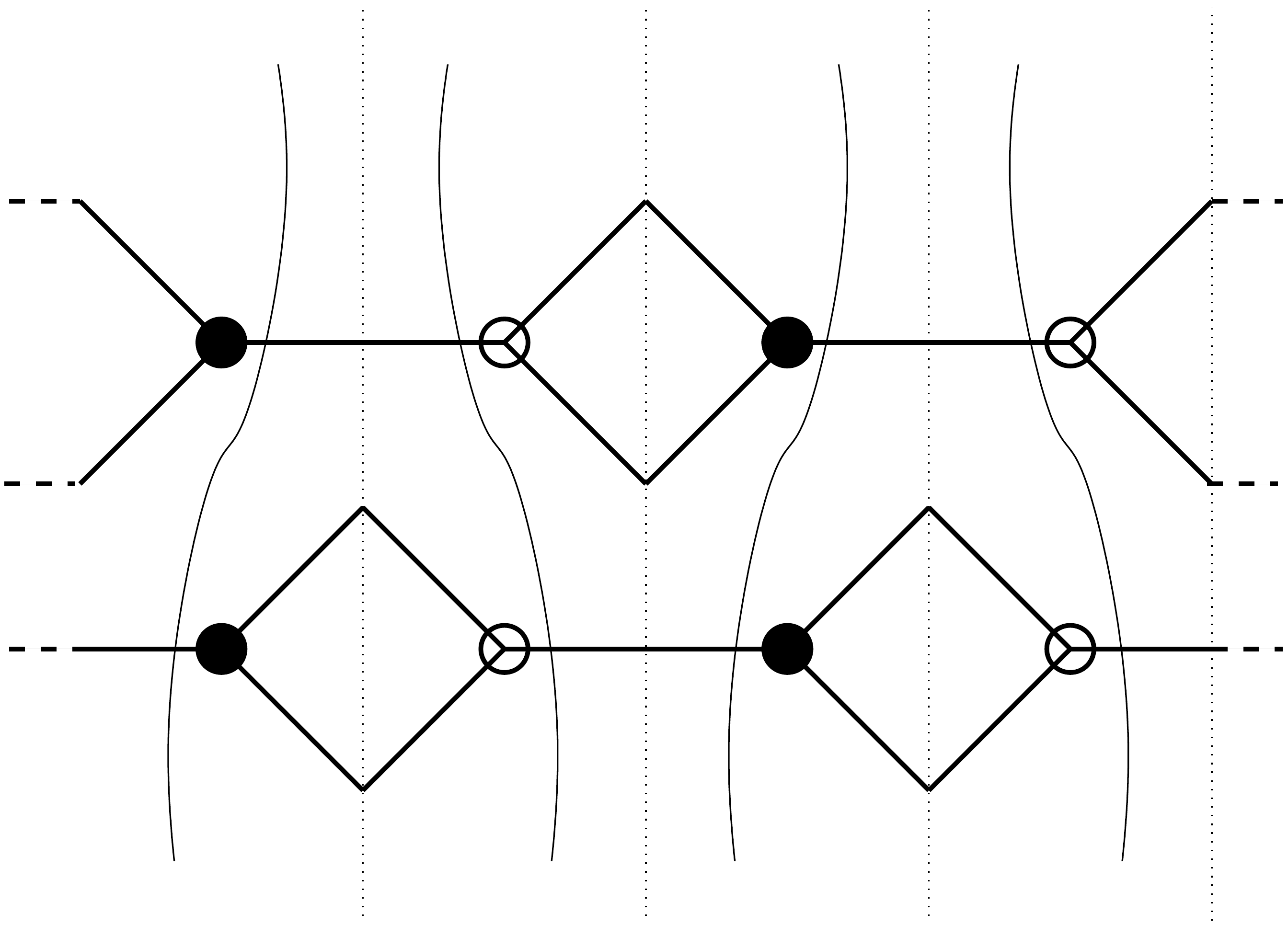}
\caption{Example of the pairing defined in lemma \ref{lblemma}.  The network is the same as in Fig.~\ref{figLdaggerL2}.  The thin curving vertical lines represent minimum cuts of the
network.  There are four such lines, each cutting one of the cases $\sigma=1,2,3,4$.  The lines for odd $\sigma$ are reflected in the horizontal direction compared to those for even
$\sigma$.  The vertical dotted lines represent reflection planes: reflecting the region between any two neighboring vertical curved lines about one of these vertical dashed lines leaves the region unchanged except for interchanging tensors $\cT$ and $\overline \cT$ (we have drawn one of these lines at the right-hand side of the figure, indicating a reflection
relating vertices at the right-hand side to those at the left-hand side).  The pairing in lemma \ref{lblemma} pairs vertices related by such a reflection.}
\label{figcutpair}.
\end{figure}

We now show the lower bound
\begin{lemma}
\label{lblemma}
Let linear operator $L$ correspond to a tensor network $\cN$ obtained from a graph $G$ with $|V|$ vertices and $|E|$ edges.  Then,
for $\scN({\rm tr}\Bigl((L^\dagger L)^k\Bigr))$ for $k\geq 1$ we have
\be
C_{\rm max}\geq k|E|-(k-1)MC(G).
\ee
\begin{proof}
The case $k=1$ is already given above.  So, assume $k>1$.

Suppose the lemma does not hold, so that $C_{\rm max}<k|E|-(k-1)MC(G)$ and hence $C_{\rm max}\leq k|E|-(k-1)MC(G)-1$.  Then from lemma \ref{sqlemma}, for sufficiently large $N$ we would have
\be
\frac{E[{\rm tr}(L^\dagger L)]^k}{E[{\rm tr}\Bigl((L^\dagger L)^k\Bigr)]} \geq c \cdot N^{(k-1)MC(G)+1},
\ee
for some positive constant $c$ (the ratio of expectation values would asymptotically tend to $1/n_{\rm max}$, where here $n_{\rm max}$ refers to the number
of pairings with maximal $C(\pi)$ for tensor network $\scN({\rm tr}\Bigl((L^\dagger L)^k\Bigr))$.
Then, from lemma \ref{exists}, for some choice of tensor $\cT_0$ the corresponding linear operator $L_0$ obeys $\rnk(L_0) \geq c \cdot N^{MC(G)+1/(k-1)}$, which is asymptotically larger than
$QMC(G,N)$, contradicting the fact that $\rnk(L) \leq QMC(G,N)$.

In addition to this proof, we give an alternative proof by
 explicitly giving a pairing $\pi$ for which $C(\pi)=k|E|-(k-1)MC(G)$.  We exemplify this pairing for a particular network in Fig.~\ref{figcutpair}.
Consider a minimum cut, with corresponding sets $\overline S,\overline T$.
For $v \in \overline T$, we pair $(v;\sigma)$ with $(v;\sigma+1)$ for odd $\sigma$, while for
$v \in \overline S$ we pair $(v;\sigma)$ with $(v;\sigma+1)$ for even $\sigma$.
Then, for every edge $(v,w)$ for vertices $v,w \in \overline T\setminus T$ there are $k$ closed loops, corresponding to
edges $((v;\sigma),(w;\sigma))$ and $((w;\sigma+1),(v;\sigma+1))$ for odd $\sigma$;
similarly, for every edge $(v,w)$ for vertices $v,w \in \overline S\setminus S$ there are $k$ closed loops, 
corresponding to edges $((v;\sigma),(w;\sigma))$ and $((w;\sigma+1),(v;\sigma+1))$ for even $\sigma$.
These edges $(v,w)$ for $v,w \in \overline T\setminus T$ or $v,w \in \overline S\setminus S$ are not in the cut set.
Similarly, for every edge in the output set or in the input set (assuming that the edge is not in the cut set) there are $k$ closed loops.
However, for each edge $(v,w)$ in the cut set, there is only one closed loop, corresponding to edges
$((v;1),(w;1)),((w;2),(v;2)), \ldots, ((w;2k),(v;2k))$.

Thus, the number of closed loops is equal to $k$ times the number of edges not in the cut set, plus the number of edges in the cut set, which
equals $k|E|$ minus $k-1$ times the number of edges in the cut set.
\end{proof}
\end{lemma}

\section{Upper Bound For Higher Moments and Its Realization By ``Direct Pairings"}
\label{mubsec}
The pairing in the lemma \ref{lblemma} has a certain structure, which we now define (that pairing is not the only pairing consistent with this structure).
\begin{definition}
\label{dsnDef}
Consider an arbitrary closed tensor network, $\scN$, with some vertices having tensor $\cT$ and some having tensor $\overline \cT$.
Suppose that there are two vertices, $v,w$ with $v$ having tensor $\cT$ and $w$ having tensor $\overline \cT$.
Further, suppose that there is at least one edge from $v$ to $w$ which has the property that the local ordering of indices makes that
edge correspond to the same index for tensor $\cT$ as it does for tensor $\overline \cT$.
Then, define a new closed tensor network, $\scN'$ by removing the vertices $v,w$ from $\scN$; every edge from $v$ to $w$ is removed, while for every pair of edges $(u,v)$ and $(w,x)$ for which the local ordering gives the same tensor index at $v$ as at $w$, we replace that pair with a single edge $(u,x)$, defining the local ordering in the obvious way, so that
the edge $(u,x)$ corresponds to the index of the tensor at $u$ that $(u,v)$ did and corresponds to the index of the tensor at $x$ that $(w,x)$ did.
Then, we say that a network $\scN'$ constructed in this fashion is a ``one-step direct subnetwork" of $\scN$; more specifically, we call it a
``one-step direct subnetwork made by pairing $v,w$ in $\cN$" to indicate how it is constructed.
Any network $\scN'$ constructed by zero or more steps of the above procedure is termed a ``direct subnetwork" of $\scN$.

Given a sequence of direct subnetworks, $\scN(1)$ pairing $(v(1),w(1))$ in $\scN$ and $\scN(2)$ pairing $(v(2),w(2))$ in $\scN(1)$, and so on, we define a partial pairing of the original network
$\scN$.  This partial pairing pairs $v(1)$ with $w(1)$, pairs $v(2)$ with $w(2)$, and so on.  This is a partial pairing as it may pair only a subset of the vertices.  Such a partial pairing is termed a direct partial pairing.  If all vertices are paired (so that the last direct subnetwork in the sequence has no vertices), then this is termed a direct pairing.

For each direct partial pairing there are several possible sequences of direct subnetworks $\scN(1),\ldots$, but the last network in the
sequence is unique, and we say that this is the direct subnetwork determined by that direct partial pairing.
\end{definition}

One way to understand direct pairings is to consider the case that the graph has degree $d=2$ and has one input and one output edge.  Then, if the graph has a single vertex, the problem of the singular values of $L$ reduces to a well-studied problem in random matrix theory, studying the singular values of a random square matrix with independent entries.  This is the so-called chiral Gaussian Unitary Ensemble\cite{chgue1,chgue2}.
In this case, it is well-known that the dominant diagrams in the large $N$ limit for any moment are the so-called ``rainbow diagrams" (these are also called ``planar diagrams")\cite{rainbow}.
Even if there is one input and one output edge, but more than one vertex (so that we now study the singular values of a power of a random matrix) the dominant diagrams are still rainbow diagrams.  These rainbow diagrams are precisely the direct pairings in this case.
If we still stick to the case $d=2$, with $|S|=|T|=MC(G)$, but allow $|MC(G)|>1$,
dominant diagrams can still be understood as rainbow diagrams: for each of the $MC(G)$ distinct paths from input to output, we draw a rainbow diagram, pairing only vertices which are both in the same path (i.e., given a power $k\geq 1$ so that we have $k$ copies of a given path with tensors $\cT$ and $k$ copies with tensors $\overline\cT$, we pair vertices between copies of that path, but only within a given path, not between paths).
Again, these are the direct pairings.

\begin{figure}
\includegraphics[width=2.0in]{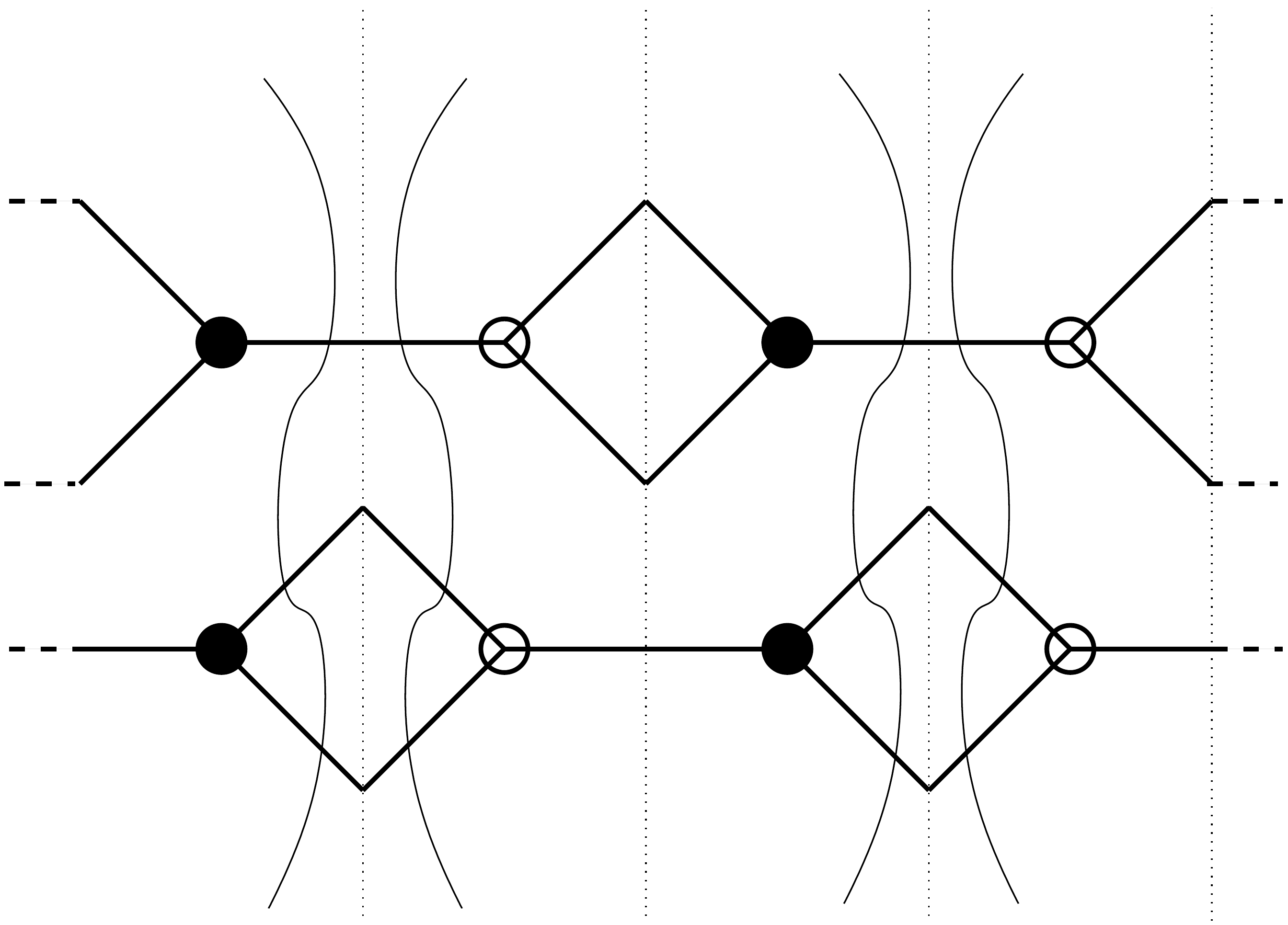}
\caption{Example of a direct pairing that does not have maximal $C(\pi)$.  The notation is the same as in Fig.~\ref{figcutpair}, except that the thin curving vertical lines
are cuts which are {\it not} min cuts.}
\label{figbadcutpair}
\end{figure}

Later, we will use this understanding in terms of rainbow diagrams to better understand the case $d>2$.  Suppose $d>2$ and we have a direct pairing.
We can construct $MC(G)$ edge disjoint paths from input to output.  We will show that the direct pairing has the properties that if we consider only the
vertices and edges in one of these paths, the result is a rainbow diagram.  Since the edge-disjoint paths might share vertices (if $d\geq 4$) this can
impose some relationship between the different rainbow diagrams for each paths: i.e., it is not the case that we can choose a distinct rainbow diagram for each path independently.  Further, while every pairing that gives rainbow diagrams when restricted to each of these paths will be a direct pairing, not every such direct pairing
will having maximal $C(\pi)$; see Fig.~\ref{figbadcutpair}.

\begin{definition}
Consider an arbitrary closed tensor network $\scN$, and a one-step direct subnetwork $\scN'$ made by pairing
vertices $v,w$ in $\scN$.
Let $\pi'$ be a partial pairing of $\scN'$.  Then, we say that $\pi'$ induces a partial pairing of $\scN$ which pairs $v$ with $w$ and pairs all other
vertices as they are paired in $\pi'$ (every vertex in $\scN$ other than $v,w$ corresponds to a vertex in $\scN'$).
Inductively, for any direct partial pairing $\theta$ defining a direct subnetwork $\scN'$ and any pairing $\pi'$ of $\scN'$
we define a partial pairing on $\scN$ induced by $\pi'$.  We write $\Pi(\pi',\theta)$ to denote this partial pairing.
\end{definition}

\begin{definition}
Given a partial pairing $\pi$ of a network, we define $C(\pi)$ the number of closed loops created by the partial pairing in the obvious way: it is the
number of distinct closed loops, where as in Definition \ref{cldef} each closed loop contains edges $(v_1,w_1),\ldots,(v_l,w_l)$ such that $w_i$ is paired with $v_{i+1}$ (identiyfing $v_{l+1}$ with $v_1$) with the local ordering assigning the same index of the
tensor to the edge $(v_i,w_i)$  at $w_i$.  Note that for a partial pairing, some edges might not be in a closed loop.
\end{definition}

\begin{lemma}
\label{create}
Consider direct partial pairing $\theta$ determining direct subnetwork $\scN'$ and let $\pi'$ be a pairing of $\scN'$.
Then
\be
C(\Pi(\pi',\theta))=C(\pi')+C(\theta).
\ee
In the special case that $\scN'$ is a one-step direct subnetwork,
$C(\theta) \leq N_E(v,w)$,
where $N_E(v,w)$ is the number of edges from $v$ to $w$, with $v,w$ as in definition \ref{dsnDef}; this is an equality if the ordering is such that all of these edges correspond to the same index at $v$ 
as they do at $w$.
\end{lemma}
When applying this lemma \ref{create}, we will say below that the pairing of vertices $v,w$ ``creates $N_E(v,w)$ closed loops" or that the pairing $\theta$ ``creates $C(\theta)$ closed loops".

\begin{lemma}
\label{ndsp}
Let $\pi$ be a pairing of a closed tensor network $\scN$.  Assume that there is no one-step direct subnetwork $\scN'$ with pairing $\pi'$ such that $\pi$ is induced by $\pi'$.
Then,
\be
C(\pi)\leq N_E(\scN)/2,
\ee
where $N_E(\scN)$ is the number of edges in tensor network $\scN$.
\begin{proof}
Every closed loop for pairing $\pi$ must be composed of at least two edges (if it is composed of one edge, then pairing those vertices defines a one-step direct subnetwork).
\end{proof}
\end{lemma}

\begin{lemma}
\label{continue}
Let $\scN=\scN({\rm tr}\Bigl((L^\dagger L)^k\Bigr))$.  Let $G$ be the graph corresponding to this tensor network.  Let $\scN'$ be a direct subnetwork of $\scN$.
Let $\pi'$ be the partial pairing defined by $\scN'$ (if $\scN'$ consists only of closed loops, then $\pi'$ is a pairing).
Then, labelling the vertices of $\scN$ by pairs $(v;\sigma)$ with $\sigma\in \{1,...,2k\}$ as above, the partial
pairing $\pi$ only pairs $(v;\sigma)$ with $(w;\tau)$ for $v=w$ and $\sigma$ odd and $\tau$ even.

Further, (*) if a pair of vertices $(x;\mu)$ and $(y;\nu)$ in $\scN'$ are connected by an edge in $\scN'$, then either $x=y$ and $\mu \neq \nu \mod 2$
and the local ordering is such that the edge is assigned the same index of the tensor at $(x;\mu)$ as it is at $(y;\nu)$,
or $\mu=\nu \mod 2$ and there is an edge $(x,y)$ in $G$ and the local orderings agree: if the ordering assigns the $j$-th index of the
tensor to edge $(x,y)$ at $x$ it also assigns the $j$-th index of the tensor to the edge $((x;\mu),(y;\nu))$ at $(x;\mu)$ and similarly for $y$ and $(y;\nu)$.
\begin{proof}
The fact that $\sigma;\tau$ have different parity mod $2$ follows because the odd and even vertices correspond to $\cT^\dagger$,$\cT$ respectively.

The statement (*)  in the last paragraph of the claim of the lemma can be established inductively.
The base case is that $\scN'=\scN$, where (*) follows trivially.  
If $\scN'$ is a direct subnetwork obeying(*) and $\scN''$ is a one-step direct subnetwork, one can check case-by-case that $\scN''$ obeys (*).

Once (*) is established, the fact that the pairing only pairs $(v;\sigma)$ with $(w;\tau)$ for $v=w$ and $\sigma$ odd and $\tau$ even follows inductively since
one only pairs vertices connected by an edge.
\end{proof}
\end{lemma}

\begin{figure}
\includegraphics[width=2.0in]{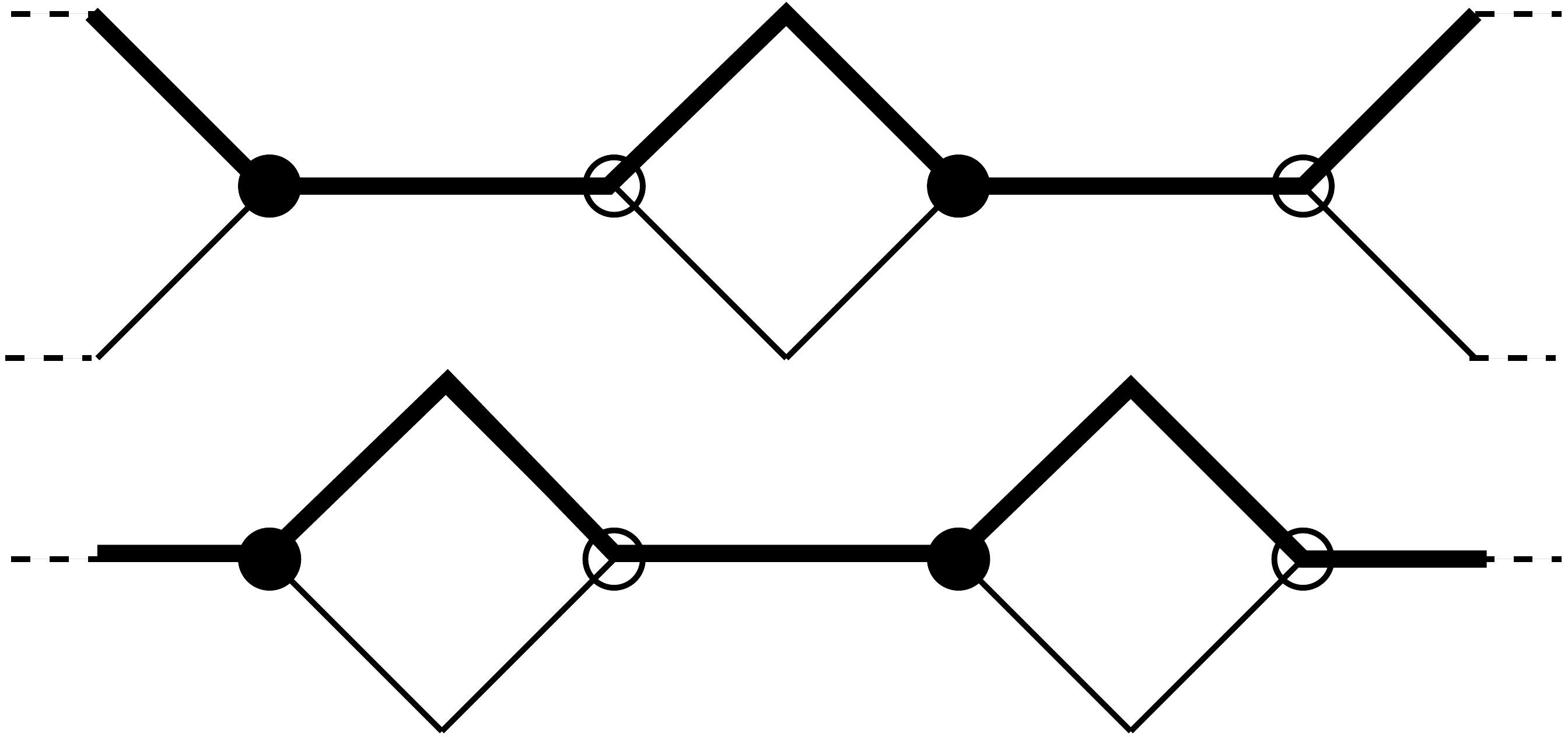}
\caption{Example of paths $Q(i)$ as constructed in lemma \ref{ublemma}.  The two thickened solid lines each represent such a path. For this network, the choice of paths
$P(i)$ is non-unique.}
\label{figQi}
\end{figure}

\begin{lemma}
\label{ublemma}
Let $\scN=\scN({\rm tr}\Bigl((L^\dagger L)^{k}\Bigr)  )$.
Then, for any direct pairing $\pi$,
\be
C(\pi)\leq  k |E|-(k-1)MC(G).
\ee
For any pairing $\pi$ that is not a direct pairing,
\be
C(\pi)\leq  k |E|-(k-1)MC(G) -1.
\ee

Finally, for any graph $G$, $n_{\rm max}$ is bounded by an exponentially growing function of $k$.
\begin{proof}
First consider the case that $\pi$ is a direct pairing.
Consider a maximal flow on the graph $G$, giving each edge of the graph capacity $1$.  By the max-flow/min-cut theorem, the flow is equal to the min cut\cite{ct1,ct2}.
Further, the flow on each edge in a maximal flow can be chosen to be an integer, and hence equals $0$ or $1$ on every edge.  Consider the set of edges
on which the flow equals $1$.  This set defines $MC(G)$ edge-disjoint paths from input to output.  Call these paths $P(1),P(2),\ldots,P(MC(G))$.

Let $v_{i,1},v_{i,2},\ldots,v_{i,l(i)}$ denote the sequence of vertices in $P(i)$, as the path is traversed from input to output, where $l(i)$ is the total
number of vertices in the path $P(i)$.
Each path in the graph defines a closed path $Q(i)$ in the network: the path is traversed backwards for each network corresponding to $L^\dagger$ and forward for each network corresponding to $L$ so that it forms a closed path traversing $kl(i)$ vertices; i.e., the path $Q(i)$ traverses vertices
$(v_{i,1};1), (v_{i,2};1),  \ldots, (v_{i,l(i)};1), (v_{i,l(i)};2),  \ldots, (v_{i,2};2), (v_{i,1};2), \ldots$.  See Fig.~\ref{figQi} for an example.
If $P(i)$ has $e_i$ edges in the graph, including $1$ input edge and $1$ output edge, then the number of edges $E_i$ in $Q(i)$ is equal to $2k (e_i-1)$.

  By lemma \ref{continue}, for a direct pairing, for every such path, vertex $(v_{i,b};\sigma)$ is paired with $(v_{i,b};\sigma')$ for some $\sigma'$.
So, $\pi$ pairs vertices in $Q(i)$ with other vertices in $Q(i)$.
As noted above, in the case of a graph with degree $d=2$, a direct pairing gives a rainbow diagram.  Here, if we consider the subgraph containing only
vertices in $Q(i)$, we have a graph with degree $2$ and again
 pairing $\pi$ 
defines a ``rainbow diagram".  Hence, the number of closed loops formed by edges on the path is equal to
$$\frac{1}{2}E_i+1=k e_i-(k-1).$$  (One can derive this number of closed loops using the known result for rainbow diagrams, or one can also derive it inductively: given a graph with degree $2$ containing $1$ closed loop with $e_i>2$ edges, pairing two vertices $v,w$ gives a one-step direct subnetwork
with $2$ fewer edges and creates $N_E(v,w)=1$ edges between them, while for $e_i=2$, the pairing gives two closed loops.)
Hence, the total number of closed loops formed in all such paths in the network is equal to $$k\sum_{i=1}^{MC(G)} e_i -(k-1) MC(G).$$

Let $Q_E$ be the set of edges in $\scN$ which are in some path $Q(i)$ and $Q_V$ be the set of vertices in $\scN$ which are in some path $Q(i)$.
We now count the number of closed loops which do not contain any edges in $Q_E$; note that for a direct pairing, every closed loop either contains only
edges from $Q_E$ or contains no edge in $Q_E$.
Let $P^\perp$ be the set of edges in $G$ which are not in a path $P(i)$, so $|P^\perp|=|E|=\sum_i e_i$.  Note that $S-MC(G)$ of the edges in $P^\perp$ are input edges and $T-MC(G)$ of such edges
are output edges so that there are $|P^\perp|-(S-MC(G))-(T-MC(G))$ edges in $G$ which are not in a path $P(i)$ and which are closed edges.
Every edge in $P^\perp$ which is closed corresponds to $2k$ edges in the network, while every edge in $P^\perp$ which is open corresponds to only $k$
edges in the network. 
The number of closed loops which can be formed by these edges is at most $1/2$ the number of edges which connect $\cT$ to $\cT$ (i.e., corresponding to the closed edges in $G$) or $\overline \cT$ to $\overline \cT$ (i.e., also corresponding to closed edges), plus the number of edges connecting $\cT$ to $\overline \cT$  (i.e., corresponding to the open edges in $G$) so that the total number of such closed loops
is at most
$(1/2) (2k) (|P^\perp|-(S-MC(G))-(T-MC(G)))+k (S-MC(G))+(T-MC(G))=k |P^\perp|$.
Hence, the total number of closed loops including edges in $Q_E$ and edges not in $Q_E$ is at most
\be
\sum_i e_i+(k-1) MC(G)+k|P^\perp|=k|E|-(k-1) MC(G).
\ee

Now, suppose that $\pi$ is not a direct pairing.  We show that $C(\pi)\leq  k |E|-(k_i-1)MC(G) -1$.
Before giving the proof, we give some motivation.  The basic idea is that if we consider the edges not in $Q_E$ then no pairing can do better than the
direct pairing: the direct pairing has one loop of length $1$ for each edge in $Q_E$ corresponding to an output edge in $G$ and has one loop of
length $2$ for each edge in $Q_E$ not corresponding to an output edge in $G$, and that is the shortest such a loop can be.  On the other hand, when
we consider the edges in $Q_E$, we know that a rainbow diagram gives the optimal pairing for a network with degree $d=2$ and so any other pairing must be
worse.
To do this in detail, we will use lemma \ref{ndsp}.

The $\theta$ be a direct partial pairing of $\scN$ determining direct subnetwork $\scN'$ such that $\pi$ is induced by a pairing $\pi'$ of $\scN'$ (i.e., $\pi=\Pi(\pi',\theta)$) and such that there is
no one-step direct subnetwork $\scN''$ of $\scN'$ with pairing $\pi''$ such that $\pi'$ is induced by $\pi''$.
By lemma \ref{ndsp}, 
$C(\pi')\leq N_E(\scN')/2$, so
by lemma \ref{create}, $C(\pi)\leq N_E(\scN')/2+C(\theta)$.
We wish to estimate $C(\theta)$ and to estimate $N_E(\scN)-N_E(\scN')$.  Recall that when we define a one-step direct subnetwork, the number of
edges changes for two reasons: we remove $N_E(v,w)$ edges (and create $N_E(v,w)$ loops) but we also combine other pairs of edges into a single edge.

Let $C$ be the set of these closed loops created by $\theta$.
Let $C_E$ be the set of edges in a loop in $C$.
Each closed loop in $C$ contains either only edges in $Q_E$ or contains no edges in $Q_E$.
Consider the loops in $C$ containing edges not in $Q_E$.  Let $R_o$ be the set of edges in these loops which correspond to output edges in $G$ and
let $R_i$ be the set of edges in these loops which do not correspond to output edges in $G$ so that the number of loops in $C$ containing edges
not in $Q_E$ is bounded by $|R_o|+(1/2) |R_i|$.  f

Consider the loops in $C$ containing edges in $Q_E$.  Each such loop contains edges from at most one path $Q(i)$.   
In such a path, there are $2k (e_i-1)$ total edges in $\scN$.  For each path $Q(i)$ we define a path in $\scN'$ in the obvious way, so that the path in $\scN'$ consists of the vertices in $\scN'$ which are
in the path in $\scN$.
If all edges in $Q(i)$ are in a loop in $C$ then there are $k(e_i-1)+1$ closed loops in $C$ containing edges in $Q(i)$.
On the other hand, if not all edges in $Q(i)$ are in such a loop, then there are less than $k(e_i-1)+1$ closed loops and the number of closed loops is equal to
$(1/2)(E_i-E'_i)$, where $E_i$ is the number of edges in $Q(i)$ and $E'_i$ is the number of edges in the corresponding path in $\scN'$.
So,
\be
N_E(\scN)-N_E(\scN')\geq |R_o|+|R_i|+\sum_i (E_i-E'_i).
\ee
Let $q$ be the number of paths $Q(i)$ such that $\pi'$ pairs all vertices in $Q(i)$.
So,
\begin{eqnarray}
C(\theta) &=& |R_o|+(1/2) |R_i|+(1/2)\sum_i (E_i-E'_i)+q \\ \nonumber
&\leq & \frac{N_E(\scN)-N_E(\scN')}{2}+|R_o|/2+q.
\end{eqnarray}
So,
\be
N_E(\scN')/2+C(\theta) \leq N_E(\scN)/2 + |R_o|/2 + q.
\ee
The number of open edges in $G$ is $|S|+|T|$ of which $|S|+|T|-2MC(G)$ edges are not in a path $Q(i)$, so
$|R_o|\leq k(|S|+|T|-2MC(G))$.
So,
\begin{eqnarray}
N_E(\scN')/2+C(\theta) &\leq& N_E(\scN)/2 +k(|S+|T|)/2-kMC(G) + q \\ \nonumber
&=& k|E|-kMC(G)+q.
\end{eqnarray}
So, unless $q=MC(G)$, we have established the desired bound on $C(\pi)$.

So, suppose that $q=MC(G)$ and suppose that $C(\pi)=C_{\rm max}$.  We will show that $\pi$ is a direct pairing.
We will use the assumption that the network is connected.  Since $C(\pi)=C_{\rm max}$, every edge in $\scN$ which is not in a loop created by $\theta$ must
either be in a loop of length $1$ (if corresponds to an open edge in $G$) or to a loop of length $2$ (otherwise).
Consider a vertex $(v;\sigma)$ which is not paired by $\theta$
and which is attached to an open edge. Since this edge must be in a closed loop of length $1$, $(v;\sigma)$ must be paired with $\sigma \pm 1$ depending on $\sigma \,{\rm mod} \, 2$ and on whether it is an input edge or an output edge.  Thus, we can pair $(v;\sigma)$ with $(v;\sigma \pm 1)$ in $\scN'$ giving a further one-step direct subnetwork $\scN''$ such that a pairing $\pi''$ on $\scN''$ induces $\pi'$.  So, we can assume that no such vertices exist.
Consider instead a vertex $(v;\sigma)$ which is not paired by $\theta$
and which neighbors a vertex $(w;\sigma)$ which is paired by $\theta$.  The pairing $\theta$ is a direct partial pairing so it pairs $(w;\sigma)$ with $(w;\tau)$ for some $\tau$.  Then, since the edge
$((v;\sigma),(w;\sigma))$ must be in a loop of length $2$ in $\pi$, the pairing $\pi$ must pair $(v;\sigma),(v;\tau)$.  So, pairing these vertices would define a one-step direct
subnetwork $\scN''$.
So, there can be no vertices not paired by $\theta$ which are attached to output edges or which neighbor a vertex paired by $\theta$; so, since the network is connected, all vertices are paired by $\theta$
and $\pi$ is a direct pairing.

We now bound $n_{\rm max}$ to bound the $c(G,k)$.  Let $\pi$ be a direct pairing with $C(\pi)=C_{\rm max}$.  In each path $Q(i)$, the direct pairing must define a rainbow diagram.
Further, since every edge not in $Q_E$ is either in a loop of length $1$ (if it is an open edge) or a loop of length $2$ (otherwise), the pairing of the vertices in $Q_V$ fully determines
$\pi$.
So, we can bound $n_{\rm max}$ by bounding the number of possible pairings of the vertices in $Q_V$.  For each path $Q(i)$, the pairings of the vertices in that path
define some rainbow diagram.  If $P(i)$ has $l(i)$ vertices, than $Q(i)$ has $2k l(i)$ vertices.  There is some restriction on the pairing of these vertices, as one can only
pair vertices corresponding to $\cT$ to those corresponding to $\overline \cT$.  Ignoring this restriction to obtain an upper bound, we ask for the number
of possible rainbow diagrams pairing $2kl(i)$ vertices (this is a problem that arises in estimating the $2kl(i)$-th moment in the Gaussian Orthogonal Ensemble where one consider random real symmetric matrices).
The number of such rainbow diagrams is at most exponential\cite{rainbowbound} in $2kl(i)$ and so
the product over paths $Q(i)$ of the number of such rainbow diagrams for each path is at most exponential in $2k\sum_i l(i)\leq 2k|V|$, showing the desired result.
The number $n_{\rm max}$ may be less than the product of these for two reasons: first, not all direct pairings have $C(\pi)=C_{\rm max}$ and second, if two paths share a vertex then the imposes some relation between the pairings on those two paths.  
\end{proof}
\end{lemma}

{\it Proof of Theorem \ref{mainth}}
By lemmas \ref{lblemma},\ref{ublemma}, for $\scN=\scN({\rm tr}\Bigl((L^\dagger L)^{k}\Bigr)  )$, we have
$C_{\max}=  k |E|-(k-1)MC(G)$.
This implies Eq.~(\ref{main2}) in theorem \ref{mainth2} with $c(G,k)=n_{\rm max}$ (we show below that $c(G,k)$ does not depend on $O$; we will not need that to prove theorem \ref{mainth}).
Using Eq.~(\ref{main2}) for $k=2$ to
estimate  $E[{\rm tr}\Bigl((L^\dagger L)^{k}\Bigr)]$ and using lemma \ref{sqlemma} to estimate $E[{\rm tr}(L^\dagger L)]$,
we find that
\be
\frac{E[{\rm tr}(L^\dagger L)]^2}{E[{\rm tr}\Bigl((L^\dagger L)^2\Bigr)]}\geq \frac{1}{c(G,2)} N^{MC(G)}-\cO(1/N).
\ee
So, by lemma \ref{exists}, theorem \ref{mainth} follows.

{\it Proof of Theorem \ref{mainth2}}
We have shown Eq.~(\ref{main2}) in theorem \ref{mainth2}.  To show that Eq.~(\ref{main2ind}) holds for the independent ensemble, note that in that ensemble, the only pairings allowed are those in which we pair  $(v;\sigma)$ with $(w;\tau)$ for $v=w$.  However, by
lemma \ref{continue} all direct pairings have that property and by lemma \ref{ublemma} the only pairings $\pi$ with $C(\pi)=C_{\rm max}$ are direct pairings.
To show that $c(G,k)$ indeed does not depend on $O$, one can either note that the possible direct pairings do not depend on $O$ or one can note that
in the enbsemble in which tensors are chosen independently, one can freely change the ordering at any vertex without altering the expectation values.
The bound on the $c(G,k)$ follows follows from the bound on $n_{\rm max}$ in lemma \ref{ublemma}.

\section{Proof of Theorem \ref{strongth}}
\label{proofstrongth}
We now prove theorem \ref{strongth}.   Let $f(x)$ be some smooth function defined for $0 \leq x < \infty$ with $0 \leq f(x) \leq 1$, $f(x)=0$ for $x\geq 2$, and $f(x)=1$ for $0 \leq x\leq 1$ and let $f(x)$ decrease monotonically with increasing $x$.  Let $P_N(\epsilon)=\int f(x/\epsilon) {\rm d}\mu_N$, with $\epsilon$ chosen later.  
We choose $f(x)$ to be smooth
so that
as $N\rightarrow \infty$, $P_N(\epsilon)$ converges to some limit $Pr(\epsilon,G)=\int f(x/\epsilon) {\rm d}\mu$, where we explicitly put the dependence on $G$ in parentheses as we will deal with this probability for several different graphs below.  We have 
\be
\int f(x/\epsilon) {\rm d}\mu_N\geq 1-\frac{E[\rnk(L)]}{QMC(G,N)},
\ee
for all $\epsilon>0$.
So, $Pr(\epsilon,G) \geq \liminf_{N \rightarrow \infty} 1-\frac{E[\rnk(L)]}{QMC(G,N)}$ for all $\epsilon > 0$.
So,
if we can show that $Pr(\epsilon,G)\rightarrow 0$ as $\epsilon\rightarrow 0^+$, then 
this establishes theorem \ref{strongth}.  For a given graph $G$, let $Pr(G)$ denote the limit of $Pr(\epsilon,G)$ as $\epsilon\rightarrow 0^+$.

We now prove that $Pr(G)=0$ using induction on the number of vertices in the graph.  The base case, a graph with no vertices, obviously has $Pr(G)=0$ since all edges must be
identity edges.  Otherwise, given a general graph, we will either find a min cut which cuts it into two graphs with fewer vertices and apply the inductive assumption, or, if no such cut exists, we will prove that $Pr(G)=0$ by estimating moments.

We need the following:
\begin{lemma}
\label{prodlemma}
Let $A$ be an $N_1$-by-$N_2$ matrix and $B$ be an $N_2$-by-$N_3$ matrix.
Assume that $A$ has at least $r_A$ singular values which are greater than or equal to $\epsilon_A$ for some $\epsilon_A$.
Assume that $B$ has at least $r_B$ singular values which are greater than or equal to $\epsilon_B$ for some $\epsilon_B$.
Then,
$AB$ has at least $r_A+r_B-N_2$ singular values which are greater than or equal to $\epsilon_A \epsilon_B$.
\begin{proof}
Let $P_B$ project onto the eigenspace of $B^\dagger B$ with eigenvalue greater than or equal to $\epsilon_B^2$.  By assumption,
$P_B$ has rank at least $r_B$.
We have $$A^\dagger B^\dagger B A \geq \epsilon_B^2 A^\dagger P_B A.$$
The non-zero eigenvalues of $A^\dagger P_B A$ are the same as  the nonzero eigenvalues of $P_B A A^\dagger P_B$.
Let $P_A$ project onto the eigenspace of $A A^\dagger$ with eigenvalue greater than or equal to $\epsilon_A^2$ so that
$$P_B A A^\dagger P_B \geq \epsilon_A^2 P_B P_A P_B.$$  By assumption,
$P_B$ has rank at least $r_B$.
The operator $P_B P_A P_B$ must have at least $r_A+r_B-N_2$ eigenvalues equal to $1$ (this can be shown by using Jordan's lemma to bring both projectors $P_A,P_B$ to a block
$2$-by-$2$ form).
So, $P_B A A^\dagger P_B$ has at least $r_A+r_B-N_2$ eigenvalues greater than or equal to $\epsilon_A^2$ and so $A^\dagger B^\dagger B A$ has at least $r_A+r_B-N_2$ eigenvalues greater than or equal to $\epsilon_A^2 \epsilon_B^2$.
\end{proof}
\end{lemma}

Consider a tensor network $\cN$ and corresponding graph $G$ and linear operator $L$.  Let $C$ be a min cut.  This cut splits the tensor network into two tensor networks $\cN_1,\cN_2$ with corresponding linear
operators $L_1,L_2$ so that $L=L_2 L_1$ and splits $G$ into two graphs, $G_1,G_2$. 
Recall that $$K=\Bigl( N^{|E|-MC(G)} \Bigr)^{-1/2} L$$ and similarly $$K_1=\Bigl( N^{|E_1|-MC(G_1)}\Bigr)^{-1/2} L_1, \, K_2=\Bigl( N^{|E_2|-MC(G_2)}\Bigr)^{-1/2} L_2.$$  Since $MC(G)=MC(G_1)=MC(G_2)$ and $|E_1|+|E_2|=|E|+MC(G)$ (this holds because the $MC(G)$ edges in the cut set become both input edges of $G_2$ and output edges of $G_1$), $K=K_1 K_2$.
If both $G_1,G_2$ have at least $1$ vertex, then $G_1,G_2$ both have fewer vertices than $G$ and so by the inductive assumption, $Pr(G_1)=Pr(G_2)=0$.
\begin{lemma}
\label{l2lemma}
In this case (i.e., where $Pr(G_1)=Pr(G_2)=0$ and where the cut splitting $G$ into $G_1,G_2$ is a min cut), $Pr(G)=0$.
\begin{proof}
Since $Pr(G_1)=0$, for any $\delta>0$, there is an $\epsilon_1>0$
such that $Pr(\epsilon_1,G_1)\leq \delta/2$.   Since $\int f(\epsilon_1) {\rm d}\mu_N$ converges to $Pr(\epsilon_1,G)$, there is an $N_0$
such that for all $N \geq N_0$, the difference $|\int f(\epsilon_1) {\rm d}\mu_N-Pr(\epsilon_1,G)|$ is bounded by $\delta/2$.
So, there is an $N_0$ such that for all $N\geq N_0$, the probability that a randomly
chosen singular value of $L_1$ for a random choice of tensors will be smaller than $\epsilon_1$ is bounded by $\delta$.
So, the probability that, for a random choice of tensors in $\cN_1$, the operator $L_1$ will have more than $\sqrt{\delta} QMC(G_1,N)$
singular values smaller than $\epsilon_1$ is bounded by $\sqrt{\delta}$ for all sufficiently large $N$.

Since $Pr(G_2)=0$ also, there is an $\epsilon_2>0$ such that 
the probability that, for a random choice of tensors in $\cN_2$, the operator $L_2$ will have more than $\sqrt{\delta} QMC(G,N)$
singular values smaller than $\epsilon_2$ is bounded by $\sqrt{\delta}$ for all sufficiently large $N$.

So, with probability at least $1-2\sqrt{\delta}$, $L_1$ has at least $(1-\sqrt{\delta}) QMC(G,N)$ singular values larger than $\epsilon_1$
and $L_2$  has at least $(1-\sqrt{\delta}) QMC(G,N)$ singular values larger than $\epsilon_2$.  So by lemma \ref{prodlemma}, 
with probability at least $1-2\sqrt{\delta}$, $L$ has at least $(1-2\sqrt{\delta}) QMC(G,N)$ singular values larger than $\epsilon_1 \epsilon_2$.
Since for any $\delta>0$ such $\epsilon_1,\epsilon_2>0$ exist, the result follows.
\end{proof}
\end{lemma}

\begin{figure}
\includegraphics[width=0.5in]{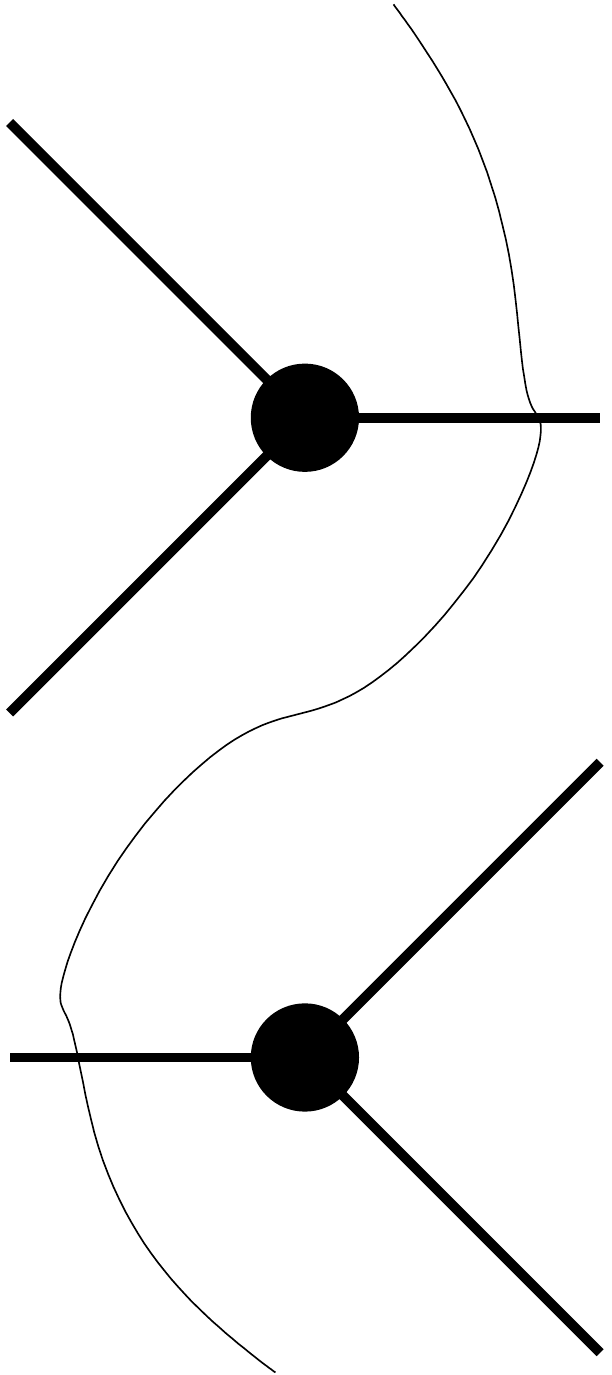}
\caption{A min cut of the network shown in Fig.~\ref{figconn}.  Thin curving vertical line shows the cut.}
\label{figconnsplit}
\end{figure}

Note that when we cut a graph, the new graph might have some open edges which connect to {\it no} vertices inside the graph.  Such edges are open edges with one open end in $S$ and one in $T$.  For example, the graph in Fig.~\ref{figconn} can be cut into two graphs, as shown in Fig.~\ref{figconnsplit}.  We call these edges ``identity edge", because the linear operator for such a graph is the identity operator (on the degree of freedom on that edge) tensored with the linear operator for the rest of the graph.
Such identity edges count as a single edge when determing $|E|$, and Eq.~(\ref{main2}) still holds for a network with identity edges; one can verify this by noting that adding an identity edge multiplies the trace of any moment of $L^\dagger L$ by $N$ and it
increases both $MC(G)$ and $|E|$
by $1$, leaving $|E|-MC(G)$ unchanged.

Suppose instead that $G$ has no min cuts other than possibly the cuts $\overline S=S, \overline T=\overline V \setminus S$ or $\overline S=\overline V \setminus T,\overline T=T$; that is, the only min cuts will cut $G$ into two graphs, one of which has no vertices.  We consider three cases: (i) $|S|=MC(G)<|T|$; (ii) $|T|=MC(G)<|S|$; (iii) $|S|=|T|=MC(G)$.
Case (i) and (ii) can be related by interchanging $L$ and $L^\dagger$ so we only do cases (i,iiii) and prove in both cases that $Pr(G)=0$.

For both cases, we will use a method of defining a new network by removing vertices from a network.
\begin{definition}
Consider a tensor network $\cN$.  We define a new network by ``removing a vertex $v$ as input from the network" as follows.  The vertex $v$ is removed.
The input edges previously attached to
$v$ are also removed.  The edges attached to $v$ which were not input edges are not removed; if they were not open edges, then they become input edges and the end attached
to $v$ becomes the open end of the edge.  If they were output edges, then they become identity edges.

Similarly, we define  ``removing a vertex $v$ as output from the network"; this definition is the same as above except ``input" and ``output" are interchanged everywhere.
\end{definition}

In case (i), we now show that the constant $c(G,k)$ in Eq.~(\ref{main2}) is equal to $1$.
Thus, in this case, the limiting distribution $\mu$ is a $\delta$-function at $1$ and so $Pr(G)=0$.

We first need the following lemma which we will also use in case (iii).  Remark: even though we are proving a combinatoric result (the value of a certain constant counting pairings), we do this in a linear algebraic fashion, by estimating
the trace of a certain product of operators.
\begin{lemma}
\label{lalemma}
Let $\cN(1),\cN(2),\ldots,\cN(l)$ be tensor networks with open edges, with $\cN(i)$ having the same number of output edges as $\cN(i+1)$ has input
edges (identifying $\scN(l+1)$ with $\scN(1)$).  Let $L_1,L_2,\ldots,L_l$ be the corresponding linear operators and let $G_1,\ldots$ be the corresponding graphs, with edge sets $E(1),\ldots$.
Consider the tensor network $\scN$ which computes the trace ${\rm tr}(L_l L_{l-1} \ldots L_1)$.
This tensor network has
\be
C_{\rm max} \leq \sum_{i=1}^l \Bigl( |E(l)|-MC(G(l))\Bigr)+{\rm min}_i MC(G(i)).
\ee
\begin{proof}
By corollary \ref{cor2}, for all $n,\epsilon_i>0$, for all sufficiently large $N$, there is a $c_i$ such that the probability that the largest singular value of $L(i)$ is
$\geq xN^{|E(i)|-MC(G(i))+\epsilon_i}$ is bounded by
$c_i x^{-n}$.  

The largest singular value of $L_l L_{l-1} \ldots L_1$ is bounded by the product over $i$ of the largest singular values of  $L_i$, so
for any $m$ there is a constant $c$ such that the probability that
 the largest singular value of $L_l L_{l-1} \ldots L_1$ is $\geq z N^{\sum_i (|E(i)|-MC(G(i))+\epsilon_i)}$ is bounded by $c z^{-m}$.  (To show this bound, one must pick each $n$ sufficiently large above)
Choosing $\epsilon_i=\epsilon/l$, it follows that
for all $\epsilon>0$, for all $m$, there is a constant $c$ such that
the probability
 the largest singular value of $L_l L_{l-1} \ldots L_1$ is 
$\geq z N^{\sum_i (|E(i)|-MC(G(i)))+\epsilon}$ is bounded by $c z^{-m}$.

The operator $L_l L_{l-1} \ldots L_1$ has rank at most ${\rm min}_i MC(G(i))$
so the 
probability that 
 the trace ${\rm tr}(L_l L_{l-1} \ldots L_1)$ is 
$\geq z N^{\sum_i (|E(i)|-MC(G(i)))+{\rm min}_i MC(G(i))+\epsilon}$ is bounded by $c z^{-m}$.
Choosing $m$ sufficiently large, this implies that
$E[{\rm tr}(L_l L_{l-1} \ldots L_1)]$ is $\cO(N^{\sum_i (|E(i)|-MC(G(i)))+{\rm min}_i MC(G(i))+\epsilon})$ for all
$\epsilon>0$.  Choosing $\epsilon<1$, this implies the result.
\end{proof}
\end{lemma}

\begin{figure}
\includegraphics[width=3in]{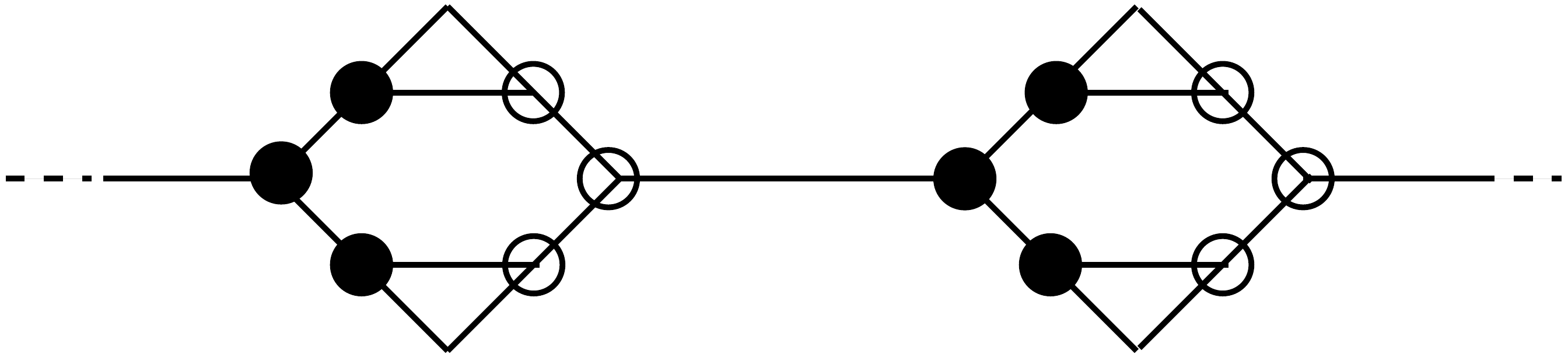}
\caption{An example network for ${\rm tr}(\Bigl(L^\dagger L\Bigr)^2)$, for a graph with $|S|=MC(G)=1$ and $|T|=4$.  For a direct pairing for which one does not pair $(v;\sigma)$ with $(v;\sigma-1)$ for any odd $\sigma$, necessarily the pairing is the same as that constructed in lemma \ref{lblemma}.}
\label{figSlessT}
\end{figure}

\begin{lemma}
\label{aBlemma}
For a graph $G$ in case (i), $c(G,k)=1$.
\begin{proof}
Let $\scN=\scN({\rm tr}(\Bigl(L^\dagger L\Bigr)^k)$.
We show that $n_{\rm max}=1$ for all $k$.  We only need to consider the case $k>1$ (for $k=1$, lemma \ref{sqlemma} shows $n_{\rm max}=1$).
Lemma \ref{lblemma} constructs a direct pairing $\pi$ with $C(\pi)=C_{\rm max}=k|E|-(k-1)MC(G)$.  There is only one min cut of the graph, so this lemma
constructs only one such direct pairing.  Let $\pi'$ be a direct pairing different from that $\pi$.
Since $\pi' \neq \pi$, there must be some vertex $(v;\sigma)$ for odd $\sigma$ which is attached to an input edge such that we pair $(v;\sigma)$ with $(v;\sigma-1)$.
For example, see Fig.~\ref{figSlessT}.

Consider the sum over all pairings $\pi''$ which pair $(v;\sigma)$ with $(v;\sigma-1)$, weighted by $N^{C(\pi'')}$.  
This is the expectation value of the one-step direct subnetwork  made by pairing
$(v;\sigma)$ with $(v;\sigma-1)$.  
Let $L$ be the linear operator defined above, and let $M$ be the linear operator corresponding to the network with vertex $v$ removed as input.

Then, the contraction of this one-step direct subnetwork is equal to ${\rm tr}(\Bigl( L^\dagger L )^{k-1} M^\dagger M\Bigr)$.  The graph $G_M$ corresponding to $M$ has
$MC(G_M) \geq MC(G)+1$ (if not, one has a min cut of $G$ with partition $\overline V = \overline S \cup \overline T$ with $\overline S=S \cup \{v\}$).
Let there be $N_E$ edges connecting $(v;\sigma)$ with $(v;\sigma-1)$ so that $G_M$ has $|E|-N_E$ edges.  Hence by lemma \ref{lalemma},
the network $\scN'$ has $C_{\rm max}\leq k|E|-(k-1) MC(G)-N_E-1$ and
so, all pairings $\pi'$ of $\scN$ which pair $(v;\sigma)$ with $(v;\sigma-1)$
have $C(\pi') \leq k|E|-(k-1)MC(G)-1$.
\end{proof}
\end{lemma}

In case (iii), we show that the constant $c(G,k)$ in Eq.~(\ref{main2}) is independent of the particular graph chosen so long as the graph has at least $1$ vertex.
In particular, the constant is the same as that if we consider a graph with $|S|=|T|=MC(G)=1$, with one vertex of degree $2$ and $|E|=2$, with
both edges open.  This case is the well-known case of random matrix theory studying the eigenvalues of $L^\dagger L$ where $L$ is a random $N$-by-$N$ matrix with independent
Gaussian entries; the matrix $L$ is drawn from the chiral Gaussian unitary ensemble.  In this case, the limiting distribution is known\cite{chgue1} and is smooth near $0$ so again $Pr(G)=0$.
\begin{lemma}
For a graph $G$ in case (iii), $c(G,k)$ is independent of $G$ so long as $G$ has at least $1$ vertex.
\begin{proof}
Let $\scN=\scN({\rm tr}\Bigl((L^\dagger L)^k\Bigr))$.
Consider the sum of all pairings $\pi$ in which for some odd $\sigma$, for some vertex $v$ attached to an input edge, $(v;\sigma)$ is paired with $(v;\sigma-1)$ and
for some vertex $w$ attached to an output edge $(w;\sigma)$ is paired with $(w;\sigma+1)$.
This is the expectation value of the  direct subnetwork $\scN'$ made by pairing
$(v;\sigma)$ with $(v;\sigma-1)$ and $(w;\sigma)$ with $(w;\sigma+1)$.  
Let $L$ be the linear operator defined above, and let $M_1$ be the linear operator corresponding to the network with vertex $v$ removed as output and let $M_2$ be the linear operator corresponding to the network with vertices $v$ removed as input and $w$ removed as output
and let $M_3$ be the linear operator corresponding to the network with vertex $w$ removed as input.
Then, the contraction of $\scN'$ equals ${\rm tr}(\Bigl( L^\dagger L )^{k-2} M_1^{\dagger} M_2 M_3^\dagger L \Bigr)$.  Let $G_{M_1},G_{M_2},G_{M_3}$ be the graphs corresponding
to $M_1,M_2,M_3$, respectively.
We have
$MC(G_{M_2}) \geq MC(G)+1$ (if not, then there is a set of $MC(G)$ edges that one can remove from $G_{M_2}$ to disconnect the graph; removing this same set of edges from $G$ will disconnect $G$ giving a min cut which cuts $G$ into two graphs, each of which has at least one vertex).

Let there be $N_E(v)$ edges connecting $(v;\sigma)$ with $(v;\sigma-1)$ 
and $N_E(w)$ edges connecting $(v;\sigma)$ with $(v;\sigma+1)$,
so $G_{M_1}$ has $|E|-N_E(v)$ edges, $G_{M_2}$ has $|E|-N_E(v)-N_E(w)$ edges, and $G_{M_3}$ has $|E|-N_E(w)$ edges.
Hence, by lemma \ref{lalemma}, $\scN'$ has $C_{\rm max}\leq  k|E|-(k-1) MC(G)-N_E(v)-N_E(w)-1$,
and so all such pairings $\pi$ of $\scN$
have $C(\pi)\leq k|E|-(k-1)MC(G)-1$.

The above proof was for $\sigma$ odd, but it works with $\sigma$ even also, if one takes the adjoint of all linear operators $L,M_1,M_2,M_3$.

So, for a maximal pairing, there is no $\sigma$ such that
for some vertex $v$ attached to an input edge, $(v;\sigma)$ is paired with $(v;\sigma-1)$ and
for some vertex $w$ attached to an output edge $(w;\sigma)$ is paired with $(w;\sigma+1)$.

For a direct pairing, there must be some $\sigma$ such that
for some vertex $v$ attached to an input edge, $(v;\sigma)$ is paired with $(v;\sigma-1)$ or
for some vertex $w$ attached to an output edge $(w;\sigma)$ is paired with $(w;\sigma+1)$.
So, for that $\sigma$, either for {\it all} $v$, $(v;\sigma)$ is paired with $(v;\sigma-1)$ are
for all {\it all} $ v$, $(v;\sigma)$ is paired with $(v;\sigma+1)$.
Suppose, without loss of generality, the first case holds, so that for all $v$, $(v;\sigma)$ is paired with $(v;\sigma-1)$ for that $\sigma$.
Define a direct subnetwork of $\scN$ by pairing such vertices $v$ in that way; this direct subnetwork is precisely the network $\scN({\rm tr}(\Bigl(L^\dagger L\Bigr)^{k-1})$, so one can apply the result above and and for some $\sigma$, either  for {\it all} $v$, $(v;\sigma)$ is paired with $(v;\sigma-1)$ are
for all {\it all} $ v$, $(v;\sigma)$ is paired with $(v;\sigma+1)$.  Iterating this $k$ times, one is left with a network with no vertices.
The possible choices (which $\sigma$ and whether one pairs with $\sigma-1$ or $\sigma+1$) are in one-to-one corresponding with rainbow diagrams and do not depend upon
$G$.
\end{proof}
\end{lemma}

\begin{figure}
\includegraphics[width=0.5in]{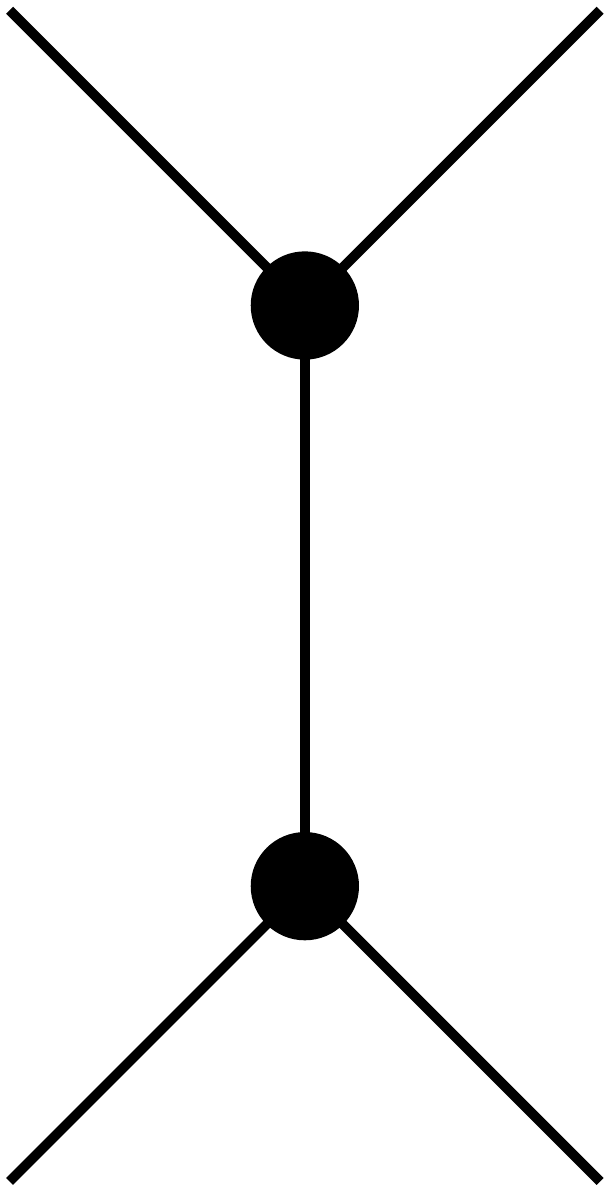}
\caption{A network with $|S|=|T|=MC(G)=2$ and with no min cuts other than those which divide $G$ into two graphs, one of which has no vertices.
Input edges are on left and output edges are on right.}
\label{fignocut}
\end{figure}

To better understand this case $|S|=|T|=MC(G)$, consider the network shown in Fig.~\ref{fignocut} as an example of such a network.
Imagine generalizing the networks that we consider, so that the $4$ open edges have capacity $N$ but the vertical edge in the figure has some capacity $N'\leq N$.
In the case $N'=1$, the tensor network factors into a product of two tensor networks; the linear operator $L$ factors as $L=L_1 \otimes L_2$, where $L_1$ maps the upper
input edge to the upper output edge and $L_2$ maps the lower input edge to the lower output edge (upper and lower refer to the position of the edge in the figure).
Then, the singular value spectrum of $L$ is the product of the singular value spectrums of $L_1,L_2$ and the limiting distribution of the singular value spectrum of $L$ is {\it not} the same as in the chiral Gaussian unitary ensemble.
(The case with $N'=1$ is equivalent to a case with degree $d=2$ but where there is a min cut which splits the graph into two networks each with only $1$ vertex; i.e., simply redraw the graph by removing the vertical edge with $N=1$.)
For general $N'$, the entries of $L$ can be obtained by summing $N'$ independent matrices of the form $L_1 \otimes L_2$.
As $N'$ increases, the correlations between entries of $L$ are reduced until eventually when $N'$ is large enough, the singular value spectrum of $L$ is the same as in the chiral Gaussian unitary ensemble.

\section{Variance}
\label{sectionvar}
Here we collect some results on variance of moments.

The first result bounds the variance
\begin{lemma}
\label{varlemma}
For any linear operator $L$ obtained from a tensor network $\scN$ with graph $G$ with $|E|$ edges, for any positive integer $c$
\be
E[{\rm tr}\Bigl((L^\dagger L)^{k}\Bigr)^c]
-E[{\rm tr}\Bigl((L^\dagger L)^{k}\Bigr)]^c=\cO(
N^{c(k |E|-(k_i-1)MC(G))-1},
\ee
and
\be
E_{\rm ind}[{\rm tr}\Bigl((L^\dagger L)^{k}\Bigr)^c]
-E_{\rm ind}[{\rm tr}\Bigl((L^\dagger L)^{k}\Bigr)]^c=\cO(
N^{2(k |E|-(k_i-1)MC(G))-1},
\ee
\begin{proof}
Let $M=L\otimes L \ldots$, with a total of
$c$ copies of $L$ in the tensor product so that
${\rm tr}\Bigl((L^\dagger L)^{k}\Bigr)^c={\rm tr}\Bigl((M^\dagger M)^k\Bigr)$.
Then, $M$ is the linear operator associated to the tensor network $\scN'$ which is the product $\scN \cdot \scN \cdot \ldots$ and has graph $G'$ with min cut $MC(G')=cMC(G)$.
We can label vertices in $G'$ by a pair $\{v;d\}$ for $d=1,2,\ldots,c$ where $d$ labelto the two different tensor factors in $M$
(we use different notation $\{v;f\}$ to distinguish this from the notation $(v;\sigma)$ used above).  Then, vertices in $\scN'$ are labelled by triples
$(\{v;d\},\sigma)$.
We can use lemma \ref{ublemma} to show that the pairings of $\scN'$ with maximal number of closed loops are direct pairings.
Every direct pairing of $\scN'$ only pairs $(\{v;d\},\sigma)$ with $(\{v;d\},\tau)$, so that the direct pairings of $\scN'$ are in one-to-one correspondance to
the product of the direct pairings of $\scN$ with itself.
\end{proof}
\end{lemma}

This has the corollary using the case $c=2$:
\begin{corollary}
With high probability, for either choice of ensemble, ${\rm av}\Bigl( (K^\dagger K)^k \Bigr)$ is within $o(1)$
of its expectation value.
\end{corollary}

To study expectation values of other products,
we define a ``product of pairings".
\begin{definition}
Let $\scN(1),\scN(2),\ldots,\scN(a)$ be a sequence of closed tensor networks, $a \geq 1$.
Let $\scN=\scN(1) \cdot \ldots \cdot \scN(a)$ be the product of these networks and label
 each vertex in the product network by a pair $[v;i]$, for $1 \leq i \leq a$, where $v$ labels a vertex in tensor network 
 $\scN(i)$. 
Let $\pi(1),\pi(2),\ldots,\pi(n)$ be pairings of these networks.  Then, define the pairing $\pi$ which is a product of these pairings
to be the pairing which pairs $[v;i]$ with $[w;i]$ is $\pi(i)$ pairs $v$ with $w$.

Note that $C(\pi)=\sum_i C(\pi(i))$.
\end{definition}

Then, once we lower bound a trace, we have an immediate lower bound for products of traces:
\begin{lemma}
\label{thisone}
Let $\scN(1),\scN(2),...\scN(a)$ be a sequence of closed tensor networks, $a \geq 1$.  Let $C_{\rm max}(i)$ be the maximum number of loops in a pairing of $\scN(i)$
and let $n_{\rm max}(i)$ be the number of distinct pairings $\pi$ of $\scN(i)$ with $C(\pi)=C_{\rm max}(i)$.  Then,
for the network $\scN$ defined to be the product network $\scN(1) \cdot \scN(2) \cdot \ldots \cdot \scN(a)$ we have
\be
\label{one}
C_{\rm max} \geq \sum_i C_{\rm max}(i),
\ee
and if $C_{\rm max}=\sum_i C_{\rm max}(i)$ then
\be
\label{two}
n_{\rm max} \geq \prod_i n_{\rm max}(i).
\ee
Further,
\be
\label{bprod}
E[\scN]\geq E[\scN(1)] \cdot E[\scN(2)] \cdot \ldots \cdot E[\scN(a)],
\ee
where $E[\scN]$ denotes the expectation value of the contraction of the network.
\begin{proof}
Let $\pi_i(j)$, for $1 \leq j \leq n_{\rm max}(i)$ label the pairings of $\scN(i)$ with $C(\pi)=C_{\rm max}(i)$.
Then, for each sequence $j_1,\ldots, j_a$, consider the product pairing of $\scN$.  This shows Eqs.~(\ref{one},\ref{two}).
To show Eq.~(\ref{bprod}), note that each expectation value $E[\scN(i)]$ is a sum over pairings of that network, weighted by the number of closed loops; the sum over product pairings
of $\scN$, weighted by the number of closed loops, is equal to $E[\scN(1)] \cdot \ldots \cdot E[\scN(a)]$.
\end{proof}
\end{lemma}

We now show:
\begin{lemma}
The expectation value of any product of traces is close to the
product of the averages:
\be
\label{prodmany}
E[{\rm tr}\Bigl( (L^\dagger L)^{k_1} \Bigr) {\rm tr}\Bigl( (L^\dagger L)^{k_2} \Bigr) \ldots]=
(1+\cO(1/N)) \cdot E[{\rm tr}\Bigl( (L^\dagger L)^{k_1} \Bigr)] E[ {\rm tr}\Bigl( (L^\dagger L)^{k_2} \Bigr)] \ldots
\ee
\begin{proof}
By lemma \ref{thisone}, $E[{\rm tr}\Bigl( (L^\dagger L)^{k_1} \Bigr) {\rm tr}\Bigl( (L^\dagger L)^{k_2} \Bigr)] \ldots \geq
(1-\cO(1/N)) \cdot E[{\rm tr}\Bigl( (L^\dagger L)^{k_1} \Bigr)] E[ {\rm tr}\Bigl( (L^\dagger L)^{k_2} \Bigr)] \ldots$.  So, we just need to
upper bound the left-hand side of Eq.~(\ref{prodmany}).

Consider the case of a product of only two traces.  Then, by Cauchy-Schwarz,
$E[{\rm tr}\Bigl( (L^\dagger L)^{k_1} \Bigr) {\rm tr}\Bigl( (L^\dagger L)^{k_2} \Bigr)] \leq
\sqrt{E[{\rm tr}\Bigl( (L^\dagger L)^{k_1} \Bigr)^2] E[ {\rm tr}\Bigl( (L^\dagger L)^{k_2} \Bigr)^2]}$.
By lemma \ref{varlemma}, and Eq.~(\ref{main2}),
\be
\sqrt{E[{\rm tr}\Bigl( (L^\dagger L)^{k} \Bigr)^2]} \leq (1+\cO(1/N)) \cdot E[{\rm tr}\Bigl( (L^\dagger L)^{k} \Bigr)].
\ee
So,
\be
E[{\rm tr}\Bigl( (L^\dagger L)^{k_1} \Bigr) {\rm tr}\Bigl( (L^\dagger L)^{k_2} \Bigr)] \leq
(1+\cO(1/N)) \cdot E[{\rm tr}\Bigl( (L^\dagger L)^{k_1} \Bigr)] E[ {\rm tr}\Bigl( (L^\dagger L)^{k_2} \Bigr)].
\ee
So, the claim follows for the case of two traces.

Now, consider a case with $a$ traces for some $a>2$.  By Cauchy-Schwarz,
\be
E[{\rm tr}\Bigl( (L^\dagger L)^{k_1} \Bigr) {\rm tr}\Bigl( (L^\dagger L)^{k_2} \Bigr) \ldots {\rm tr}\Bigl( (L^\dagger L)^{k_a} \Bigr)]
\leq 
\sqrt{E[{\rm tr}\Bigl((L^\dagger L)^{k_1}\Bigr)^2]}
\sqrt{E[{\rm tr}\Bigl( (L^\dagger L)^{k_2} \Bigr)^2 \ldots {\rm tr}\Bigl( (L^\dagger L)^{k_{a}} \Bigr)^2]}.
\ee
Applying Cauchy-Schwarz again to the second term in this product on the right-hand side, and continuing in this fashion, we find that
\be
E[{\rm tr}\Bigl( (L^\dagger L)^{k_1} \Bigr) {\rm tr}\Bigl( (L^\dagger L)^{k_2} \Bigr) \ldots {\rm tr}\Bigl( (L^\dagger L)^{k_a} \Bigr)]
\leq
\ldots
E[{\rm tr}\Bigl((L^\dagger L)^{k_1}\Bigr)^2]^{1/2}
E[{\rm tr}\Bigl((L^\dagger L)^{k_2}\Bigr)^4]^{1/4}
E[{\rm tr}\Bigl((L^\dagger L)^{k_3}\Bigr)^8]^{1/8} \ldots
\ee
By lemma \ref{varlemma}, and Eq.~(\ref{main2}),
\be
E[{\rm tr}\Bigl( (L^\dagger L)^{k_1} \Bigr) {\rm tr}\Bigl( (L^\dagger L)^{k_2} \Bigr) \ldots] \leq
(1+\cO(1/N)) \cdot E[{\rm tr}\Bigl( (L^\dagger L)^{k_1} \Bigr)] E[ {\rm tr}\Bigl( (L^\dagger L)^{k_2} \Bigr)] \ldots
\ee
\end{proof}
\end{lemma}

\section{Numerics}
\label{sectionnumerics}
\begin{figure}
\includegraphics[width=2in]{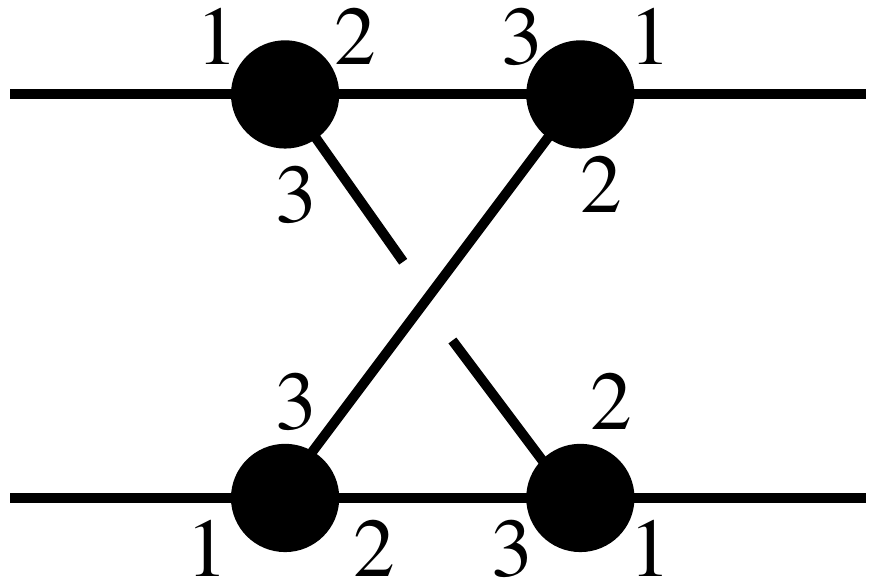}
\caption{A tensor network from Ref.~\onlinecite{mfmc2} that we studied numerically.  Numbers indicate local ordering.}
\label{fignum}
\end{figure}

\begin{figure}
\includegraphics[width=5in]{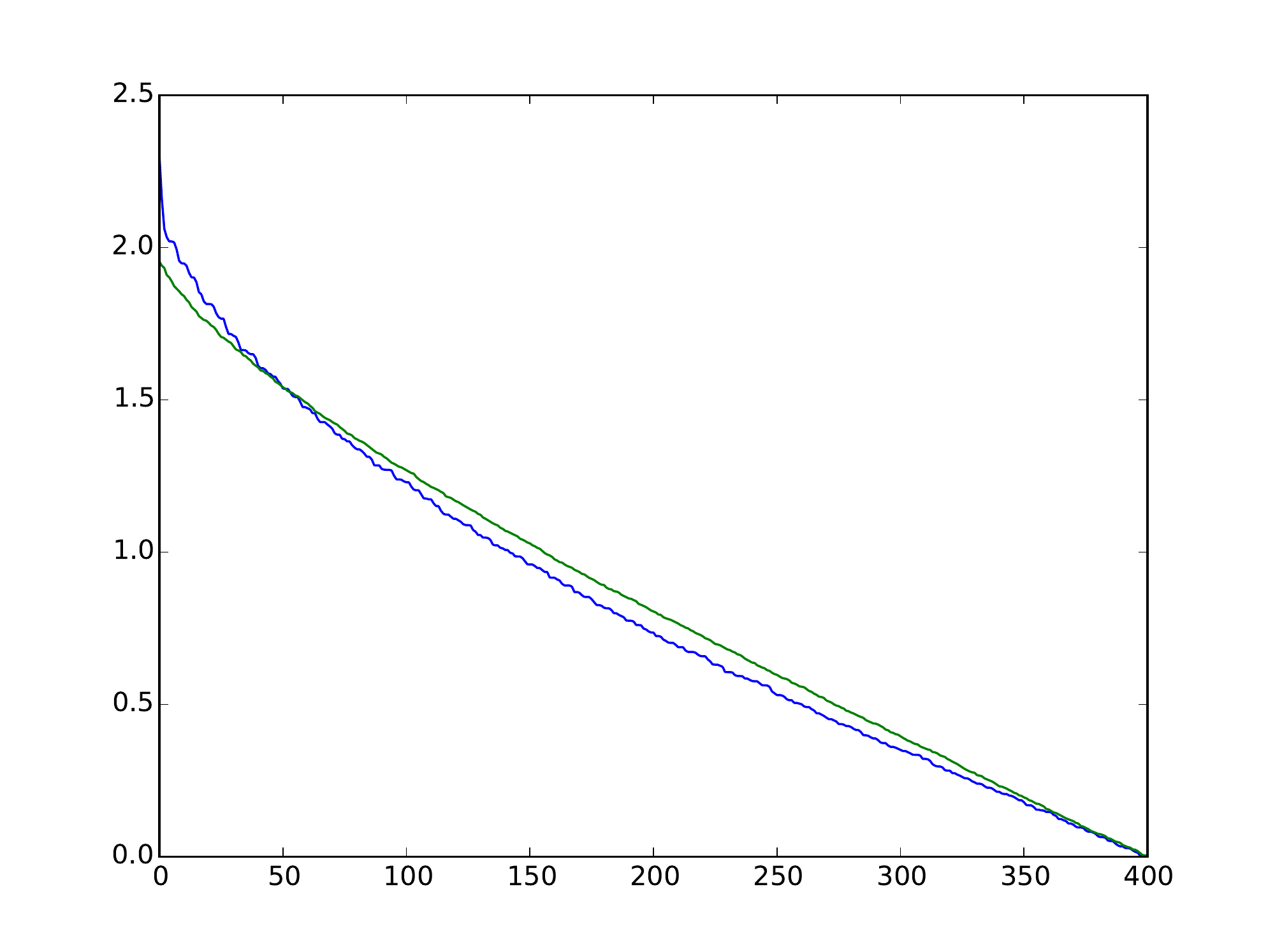}
\caption{Plot of the value of the singular values (divided by $N^3$) for one sample of the tensor network shown in Fig.~\ref{fignum} for $N=20$ shown in blue and for a chiral Gaussian unitary matrix of size $N^2$-by-$N^2$ shown in green.  The largest singular value shown on the blue curve is larger
than the largest on the green curve.}
\label{fig20}
\end{figure}

\begin{figure}
\includegraphics[width=5in]{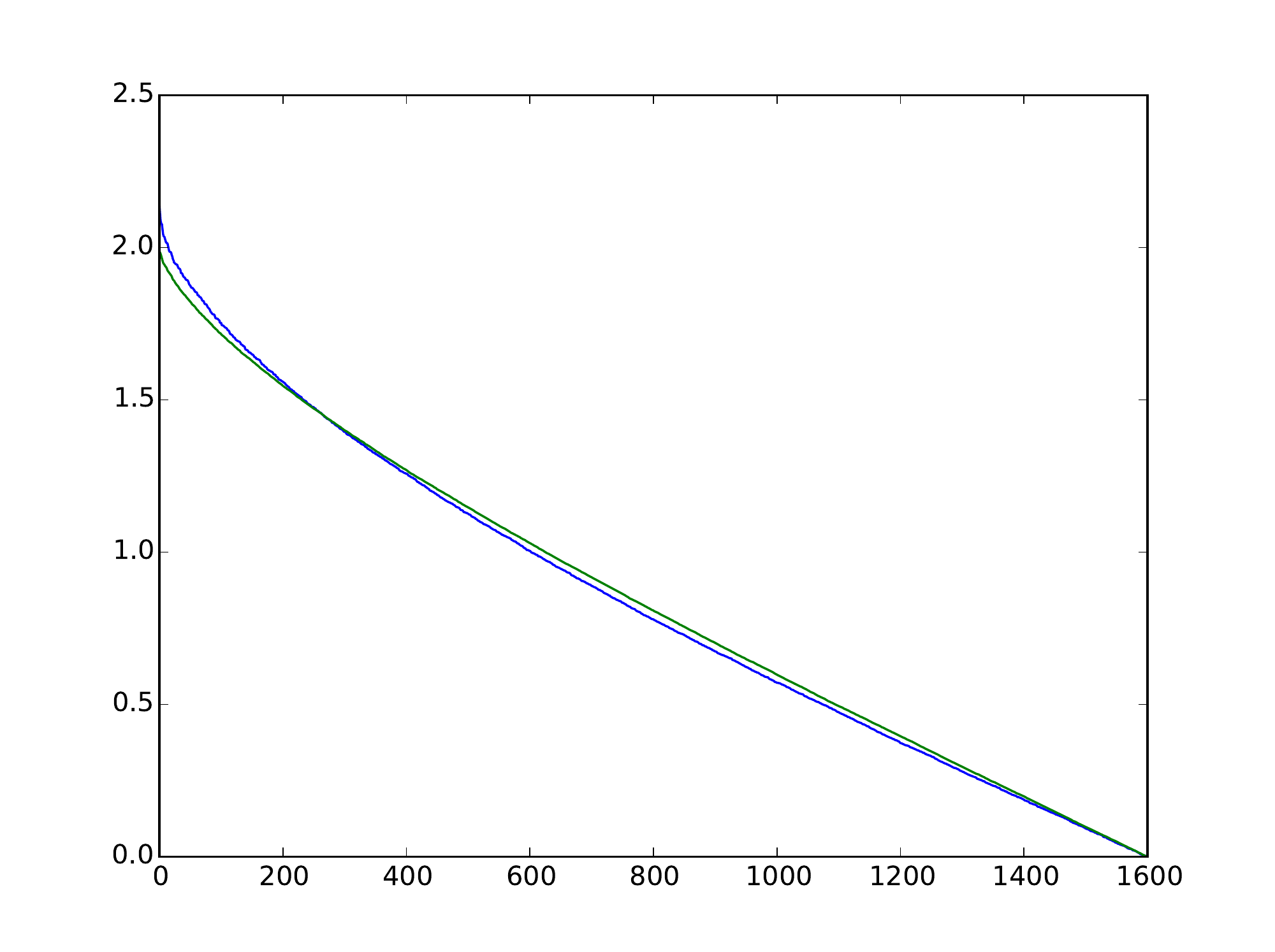}
\caption{Plot of the value of the singular values (divided by $N^3$) for one sample of the tensor network shown in Fig.~\ref{fignum} for $N=40$ shown in blue and for a chiral Gaussian unitary matrix of size $N^2$-by-$N^2$ shown in green.  The largest singular value shown on the blue curve is slightly larger
than the largest on the green curve but the discrepancy is smaller than in the case $N=20$.}
\label{fig40}
\end{figure}

We performed a numerical study of the network shown in Fig.~\ref{fignum}.  This network was used in Ref.~\onlinecite{mfmc2} as an example of a network for which
$QMF<QMC$ for $N=2$.  Since this network has $|S|=|T|=MC(G)=2$ and there is no min cut that cuts the graph into two graphs which each have at least one vertex, the results
above predict that for large $N$, the limiting distribution of singular values is, up to dividing by $\sqrt{N^{|E|-MC}}=N^3$, the same as that for random matrix theory in the chiral Gaussian unitary ensemble.
In Fig.~\ref{fig20} we show the singular values both for a randomly chosen example of this ensemble for $N=20$ and for a randomly chosen chiral Gaussian unitary matrix of size $N^2$-by-$N^2$.
In Fig.~\ref{fig40}, we show the same for $N=40$.  One can see the convergence.

We have also considered the smallest singular values to obtain numerical evidence for the rank of $L$.
Studying a few random samples for each $N$ with $2 \leq N \leq 40$, we
found that there was exactly one very small singular value (smaller than $10^{-14}$) for $N=2,3 \mod \, 4$ and no very small singular values otherwise.  In both cases, all the other
singular values were larger than $6*10^{-4}$.  While this is only numerical evidence, it supports a conjecture that for this network, $QMF(G,N,O)=QMC(G,N)-1$ if $N=2,3  \mod \, 4$ and
$QMF(G,N,O)=QMC(G,N)$ otherwise.

The quantum max-flow/min-cut conjecture of Ref.~\onlinecite{mfmc1} was shown to be false in Ref.~\onlinecite{mfmc2}.  If the conjecture above for the network shown in Fig.~\ref{fignum} is true,
then it is not even the case that ``for all $G,O$, for all sufficiently large $N$, $QMF(G,N,O)=QMC(G,N)$".  We may still hope for the weaker conjecture that
``for all $G,O$, for infinitely many $N$, $QMF(G,N,O)=QMC(G,N)$".
Indeed, to show this weaker conjecture, it suffices to show that ``for all $G,O$, for some $N_0>1$, $QMF(G,N_0,O)=QMC(G,N_0)$" as then $QMF(G,N_0^k,O)=QMC(G,N_0^k)$ for all integer $k\geq 0$ due to the following:
\begin{lemma}
For all $G$, $QMF(G,N_1 N_2,O) \geq QMF(G,N_1,O) QMF(G,N_2,O)$.
\begin{proof}
Let $\cT_1,\cT_2$ be tensors whose indices range from $1$ to $N_1$ or $1$ to $N_2$ respectively, such that the network with tensor $\cT_1$ gives a linear operator $L_1$
with rank $QMF(G,N_1,O)$ and with tensor $\cT_2$ gives a linear operator $L_2$ with rank $QMF(G,N_2,O)$.  Then, $\cT=\cT_1 \otimes \cT_2$ is a tensor whose indices
range from $1$ to $N_1 N_2$ and gives a linear operator $L=L_1 \otimes L_2$ with rank $QMF(G,N_1,O) QMF(G,N_2,O)$.  Here, the product $\cT_1 \otimes \cT_2$ is defined as follows: label the each index of $\cT$ by a pair $(i,j)$.  Write the entries of $\cT$ as $\cT_{(i_1,j_1),(i_2,j_2),\ldots,(i_d,j_d)}$.  Set $\cT_{(i_1,j_1),(i_2,j_2),\ldots,(i_d,j_d)}=(\cT_1)_{i_1,i_2,\ldots,i_d} (\cT_2)_{j_1,j_2,\ldots,j_d}$. 
\end{proof}
\end{lemma}

\appendix

\begin{thebibliography}{99}
\bibitem{mfmc1} D. Calegari, M. Freedman, and K. Walker, ``Positivity of the universal pairing in 3 dimensions",
Jour. Amer. Math. Soc. {\bf 23}, no. 1, 107-188 (2010).


\bibitem{mfmc2} S. X. Cui, M. H. Freedman, O. Sattath, R. Stong, and G. Minton, ``Quantum Max-flow/Min-cut", arXiv:1508.04644.

\bibitem{carleman} For a sequence of distributions where the moments converge to a limit $c(G,k)$ obeying Carleman's condition
$\sum_{k=1}^\infty c(G,2k)^{\frac{1}{2k}}=\infty$, the  distributions converge weakly to a limiting distribution.
See http://mathoverflow.net/questions/230794/converging-to-moments-obeying-carlemans-condition .  Sketch: since the second moment is bounded, the sequence is tight; by Prokhorov's theorem, there is a subsequence which converges to a limit $\mu$; the moments of the limiting distribution $\mu$ are the limits of the moments (to show that $\mu$ has the correct $k$-th moment, use a bound on a higher moment to establish uniform integrability) so by Carleman's condition, all subsequences which converge must converge to the same limit; so by Prokhorov's theorem, the sequence converges without needing to pass to a subsequence.
In this particular case, since the $2k$-th moments of $\mu^{\rm ind}_N$ (or $\mu_N$) are bounded by $c_1 \cdot c_2^k$ for some constants $c_1,c_2$, it is simpler.  To show that $\int f(x) {\rm d} \mu^{\rm ind}_N(x)$ has a limit as $N \rightarrow \infty$ for bounded Lipschitz functions $f(x)$, approximate $f(x)$ on an interval $[-c,+c]$ by a polynomial $p(x)$ for any $c>c_2$ and use a bound on higher moments to bound the integral $\int_{|x|>c} p(x) {\rm d} \mu^{\rm ind}_N(x)$.

\bibitem{rainbow} ``Rainbow diagrams" for random matrix theory are another term for ``planar diagrams".  This is a class of diagrams that appear in computing expectation values of traces of powers of a random matrix.  See, for example, A. Zee, {\it Quantum Field Theory in a Nutshell}, Second Edition, pp 396-400, Princeton University Press (Princeton, NJ 2010).

\bibitem{chgue1} E.V. Shuryak and J.J.M. Verbaarschot, Nucl. Phys. A {\bf 560}, 306 (1993).

\bibitem{chgue2} J.J.M. Verbaarschot, Phys. Rev. Lett. {\bf 72}, 2531 (1994); Phys. Lett. B {\bf 329}, 351 (1994).

\bibitem{rainbowbound} Since the limiting distribution in the Gaussian orthogonal ensemble is the Wigner semi-circle, which is bounded, one can deduce that the expected trace of the $k$-th moment of a matrix drawn from this ensemble is at most exponential in $k$.  See also E. Brezin, C. Itzykson, G. Parisi, and J. B. Zuber, ``Planar Diagrams", Commun. Math. Phys. {\bf 59}, 35 (1978).  Alternatively, one can reduce the problem of counting rainbow diagrams in this case to the problem of counting parenthesizations, which are bounded by an exponential.

\bibitem{ct1} P. Elias, A. Feinstein, and C. E Shannon, ``A note on the maximum
flow through a network", Information Theory, IRE Transactions on, {\bf 2}(4),117–
119 (1956).

\bibitem{ct2} L. R Ford and D. R Fulkerson, ``Maximal flow through a network",
Canadian journal of Mathematics, {\bf 8(3)}, 399–404 (1956).
\end{thebibliography}
\end{document}